Title: Basalt or not? Near-infrared spectra, surface mineralogical estimates, and meteorite analogs for 33 $V_p$-type asteroids

Short title: Basaltic asteroid survey

Authors: Paul S. Hardersen, Vishnu Reddy, Edward Cloutis, Matt Nowinski, Margaret Dievendorf, Russell M. Genet, Savan Becker, Rachel Roberts

Complete postal addresses:

Paul S. Hardersen, Planetary Science Institute, 1700 E. Fort Lowell Road, Tucson, AZ 85719-2395

Vishnu Reddy, Lunar and Planetary Laboratory, Department of Planetary Sciences, University of Arizona, 1629 E. University Boulevard, Tucson, AZ 85721-0092

Edward Cloutis, Department of Geography, University of Winnipeg, Winnipeg, MB R38 2E9

Matt C. Nowinski, 20406 Rosemallow Court, Sterling, VA 20165

Margaret Dievendorf, University of North Dakota, Department of Space Studies, 4149 University Avenue, Clifford Hall, Room 512, Grand Forks, ND 58202-9008

Russell M. Genet, 4995 Santa Margarita Lake Road, Santa Margarita, CA 93453

Savan Becker, 1218 Form Court, Odenton, MD 21113

Rachel Roberts, University of North Dakota, Department of Space Studies, 4149 University Avenue, Clifford Hall, Room 512, Grand Forks, ND 58202-9008

Corresponding author email: phardersen@psi.edu

# Abstract

Investigations of the main-asteroid belt and efforts to constrain that population's physical characteristics involve the daunting task of studying hundreds of thousands of small bodies. Taxonomic systems are routinely employed to study the large-scale nature of the asteroid belt because they utilize common observational parameters, but asteroid taxonomies only define broadly observable properties and are not compositionally diagnostic (Tholen, 1984; Carvano et al., 2010; Hasselmann et al., 2012).

This work builds upon the results of Hardersen et al. (2014, 2015), which has the goal of constraining the abundance and distribution of basaltic asteroids throughout the main asteroid belt. We report on the near-infrared (NIR: 0.7 to 2.5-μm) reflectance spectra, surface mineralogical characterizations, spectral band parameter analysis, and meteorite analogs for 33 $V_p$ asteroids. NIR reflectance spectroscopy is an effective remote sensing technique to detect most pyroxene group minerals, which are spectrally distinct with two very broad spectral absorptions at ~0.9- and ~1.9-μm (Cloutis et al., 1986; Gaffey et al., 2002; Burbine et al., 2009).

Combined with the results from Hardersen et al. (2014, 2015), we identify basaltic asteroids for ~95% (39/41) of our inner-belt $V_p$ sample, but only ~25% (2/8) of the outer-belt $V_p$ sample. Inner-belt basaltic asteroids are most likely associated with (4) Vesta and represent impact fragments ejected from previous collisions. Outer-belt $V_p$ asteroids exhibit disparate spectral, mineralogic, and meteorite analog characteristics and likely originate from diverse parent bodies. The discovery of two additional likely basaltic asteroids provides additional evidence for an outer-belt basaltic asteroid population.

Introduction. Of the many currently unanswered questions in the field of asteroid science, one significant puzzle involves the initial formation, abundance, disruption, and resultant distribution of basaltic asteroids in the main asteroid belt (1.8 AU < $a$ < 3.4 AU). Formed in the early epoch of the solar system in the chaotic regime of planetesimals, giant planet migration, and mass ejection from the solar system, the remaining basaltic asteroids in the main asteroid belt today inform us of the thermal, chemical, and temporal characteristics of the first 10s of millions of years of the solar system (Cameron, 1995; Keil, 2000; Warren, 2011).

Basalt is an extrusive, volcanic igneous rock type that forms from the high temperature melting of rocks and their resulting recrystallization (Klein and Hurlbut, 1977). This is a very common rock type on terrestrial bodies including Earth, Earth's Moon, Mercury, Venus, Mars, and asteroids (Burns, 1993). How do we know there should be basaltic material in the asteroid belt? The relatively large abundance of grouped and ungrouped iron meteorite types strongly suggest that many proto-planets chemically differentiated at high temperatures (Keil, 2000; Wasson, 1990). Compositional zoning of individual bodies into canonical crust, mantle, and core components, whether partially or fully differentiated, were later disrupted via collisions and became constituents of the solar system. However, previous researchers have noted the dearth of mantle (i.e., olivine) and basaltic NIR spectral signatures on asteroids in the present era, which could be attributed to mass loss from the solar system (Burbine et al., 1996; Sanchez et al., 2014; Brasil et al., 2017).

The terrestrial collection of basaltic achondrites themselves provide direct evidence of basalt and derivation from partial or complete planetary-scale melting (Grady, 2000; Mittlefehldt et al., 1998). The identification of the $^{26}$Al radionuclide heating mechanism and, to a lesser extent, the T Tauri induction heating method, provide pathways for this heating to planetary melting temperatures (Herbert et al., 1991; Grimm and McSween, 1993). Other types of achondrites, such as anomalous basaltic achondrites, acapulcoites, lodranites, iron meteorites (grouped and ungrouped), and pallasites also highlight even broader and more variable examples of melting processes (Mittlefehldt et al., 1998). **Most of these achondrites, the Howardite-Eucrite-Diogenite (HED) suite, are thought to derive from (4) Vesta. However, some of these achondrites, such as NWA 1240 and NWA 4587 (among others), are ungrouped and exhibit anomalous oxygen isotope ratios different from the HED meteorites, which suggest origins from parent bodies distinct from (4) Vesta (Mittlefehldt et al., 1998; Barrat et al., 2003; Connolly et al., 2007).**

The presence of (4) Vesta in the main asteroid belt as the sole remaining intact, large basaltic asteroid is additional direct evidence of that early solar system heating event. Remote investigations of Vesta across several decades and in-situ spacecraft study have confirmed Vesta's geologic character and its uniqueness within the solar system volume that encompasses the hundreds of thousands of known main-belt asteroids (McCord et al., 1970; Gaffey, 1997; Reddy et al., 2012c; Russell et al., 2012).

Evidence of the remaining basaltic asteroid population beyond (4) Vesta lies with the "Vestoids", which are posited as ejected fragments from Vesta's surface (Hardersen et al., 2014, and references therein). These Vestoids are defined using dynamical, taxonomic, mineralogic, and/or meteoritic lines of reasoning, although most potential Vestoids are still dynamically and taxonomically classified as V-type asteroids (Tholen, 1984; Bus and Binzel, 2002; Hasselmann et al., 2012; Zappala et al., 1995; Nesvorny, 2015).

In the outer asteroid belt beyond 2.5 AU, the only confirmed basaltic asteroid is (1459) Magnya, which has a semimajor axis of ~3.14 AU, is taxonomically classified as a V-type asteroid, is thought to be too far from (4) Vesta to derive from that parent body (Lazzaro et al., 2000). This asteroid also exhibits NIR spectral and mineralogical features that are consistent with a basaltic surface (Hardersen et al., 2004).

Beyond this evidence, though, our knowledge of main-belt basaltic asteroids gets decidedly murkier. Recent work has been attempting to better define the basalt that resides throughout the main belt (Vilas et al., 2000; Burbine et al., 2001; Kelley et al., 2003; Cochran et al., 2004; Duffard et al., 2006; Duffard and Roig, 2009; Moskovitz et al., 2008; Nesvorny et al., 2008; Lim et al., 2011; Mayne et al., 2011; Reddy et al., 2011; Hardersen et al., 2014, 2015; Leith et al., 2017). A larger abundance of V-type asteroids has been identified and classified, which includes a tantalizing collection of possible non-Vesta-related basaltic asteroids (Hasselmann et al., 2012; Mainzer et al. 2012). Increasing numbers of Vestoids are being characterized that have analogs within the Howardite-Eucrite-Diogenite (HED) suite of basaltic achondrites (Hardersen et al., 2014, 2015, and references therein). However, no other outer-belt basaltic asteroids have yet been identified.

The present work builds upon, and expands, the near-infrared (NIR: 0.7 to 2.5-μm) spectral and mineralogic characterization of the $V_p$ taxonomic class of asteroids (Carvano et al., 2010), which began with the work of Hardersen et al. (2014, 2015). The primary goals of this research are to: 1) identify those inner-belt $V_p$ asteroids that have a significant likelihood of having a basaltic surface composition and deriving from (4) Vesta (as Vestoids), 2) better constrain the abundance and distribution of outer-belt basaltic asteroids, and 3) test the predictive ability of the $V_p$ taxonomic class to accurately identify basaltic asteroids.

The above goals will be accomplished by: 1) visually inspecting asteroid NIR spectra and identify those that are consistent with NIR spectra of HED meteorites, 2) using a three-tiered test to determine those $V_p$ asteroids that exhibit NIR spectral and mineralogic properties consistent with a basaltic mineralogy and HED meteorite analogs, 3) constraining the surface mineralogy and potential meteorite analogs for the non-basaltic $V_p$ asteroids, and 4) combining the results from Hardersen et al. (2014, 2015) with this work to estimate the success rate of the $V_p$ taxonomy in predicting a basaltic surface mineralogy and composition.

Testing for a basaltic mineralogy involves comparing the derived ~0.9-μm (Band I) and ~1.9-μm (Band II) spectral absorption features with those of the HED meteorites, testing the consistency of the Band I vs. Band Area Ratio (BAR) plots of the $V_p$ asteroids with the basaltic achondrites (Gaffey et al., 1993, 2002), and using existing pyroxene and basaltic achondrite laboratory calibrations to estimate average surface pyroxene chemistries and comparing the results with pyroxene mineral chemistries of the HED meteorites (Gaffey et al., 2002; Burbine et al., 2009; Reddy et al., 2012a).

For those $V_p$ asteroids that exhibit Band I and Band II absorptions but are not classified as basaltic via the testing methods described above, efforts will be made to determine if the asteroids belong to a different taxonomy (e.g., S-type, M-type, etc.) and have an average surface composition that is consistent with different meteorite types. $V_p$ asteroids with featureless NIR spectra will be considered for taxonomic reclassification and efforts to constrain the potential range of meteorite types will be attempted.

The predictive test of the $V_p$ class is an important component of this work because this taxonomy uses an average five-color spectrum derived from Sloan Digital Sky Survey (SDSS) colors to define its asteroid taxonomic classes (Carvano et al., 2010). The connection between asteroid taxonomic classes, the parameters that define those classes, and the geologic nature of asteroids is not direct nor guaranteed. Compositional suggestions from an asteroid taxonomy can only be made broadly and must be done with caution (Tholen, 1984). Our sample of $V_p$ asteroids also have NEOWISE-derived albedos that are broadly consistent with (4) Vesta (Tedesco et al., 2002; Masiero et al., 2011; Mainzer et al., 2016), which together provides the best indirect evidence of a likely basaltic nature for this taxonomic group of asteroids.

$V_p$ asteroid taxonomy. The Carvano et al. (2010) asteroid classification system is based on Sloan Digital Sky Survey (SDSS) broad-band filter observations of a relatively large number of asteroids (63,468), but with that useful and large number of taxonomic classifications comes some unique characteristics. **SDSS primarily obtains photometric data of stars, galaxies, and quasars, but the data has also been applied to study asteroids (Adelman-McCarthy et al., 2006).** First, the five SDSS colors (*u'g'r'i'z'*) are solar-corrected and used to create very low-resolution, five-element, visible wavelength spectra that are the primary criterion used to define the different asteroid taxonomic classes. Second, the median spectra that define a given taxonomic class (Figure 2, Carvano et al., 2010) have associated color gradient (i.e., spectral slope) limits that define the range of acceptability of observations that fit a given taxonomic class (Figure 1, Carvano et al., 2010). The color gradient limits are the largest for the $V_p$ class for the (*i*-*r*) and (*z*-*i*) color indices (Carvano et al., 2010). In addition, the color gradient ranges for the different classes exhibit some overlap and Figure 2 from Carvano et al. (2010) shows that the $V_p$ class overlaps with the $Q_p$ class and slightly with the $O_p$ class for the (*z*-*i*) color index.

Third, SDSS observations can be of poor quality and identified as such with low taxonomic probability scores or larger than average filter errors, single observations for an asteroid may be assigned to multiple taxonomic classes, and multiple observations of individual asteroids may potentially lead to the assignment of multiple classifications for the same asteroid (Carvano et al., 2010).

**By comparison, the SDSS-based taxonomy of Roig and Gil-Hutton (2006) applied Principal Component Analysis (PCA) to identify 499 V-type, and potentially basaltic, asteroids from the SDSS Moving Object Catalog. Roig and Gil-Hutton (2006) identify 17 V-type asteroids that were also classified in Hasselmann et al. (2012) and analyzed in Hardersen et al. (2014, 2015), and this work.**

Connecting the $V_p$ taxonomy to asteroid basaltic mineralogy. For the $V_p$ taxonomy to be effective, there should be a direct connection between the primary measurable feature of the $V_p$ taxonomy and that feature's ability to securely predict an asteroid with a primarily basaltic surface composition. In this case, the primary $V_p$ taxonomic measurable is the deep absorption in the ~0.9-μm spectral region based on the solar-corrected SDSS color magnitudes at the *i'* and *z'* filters where the reflectance at the *z'* filter decreases more than any other Carvano et al. (2010) taxonomic class. The median log reflectance spectrum of the $V_p$ class is given in Figure 2 of Carvano et al. (2010), but it is not possible to strongly constrain the absolute depth (based on visual examination, we estimate an *i'* to *z'* filter decrease in reflectance of ~20%). Also, the band minimum of this median absorption feature cannot be determined due to the long-wavelength cutoff of the *z'* filter central wavelength at 0.913-μm, although the z' band goes out to 1.123-μm (Fukugita et al., 1996; Carvano et al., 2010).

In addition, the absorption feature(s) for the $V_p$ taxonomy should be caused by the major mafic silicate and spectrally active mineral(s) found in the Howardite-Eucrite-Diogenite (HED) and basaltic achondrite meteorites, which is low-Ca pyroxene (i.e., pigeonite) (Gaffey, 1976; Rubin, 1997; Mittlefehldt et al., 1998). The other abundant mafic silicate mineral that is spectrally present in the ~1-μm region is olivine (Burns, 1993), as well as olivine-pyroxene mineral mixtures (Singer, 1981; Cloutis et al., 1986; Gaffey et al., 1993). The presumption is that the $V_p$ class with its very deep absorption is a proxy for the typically deep ~0.9-μm low-Ca or orthopyroxene absorption seen in the near-infrared (NIR) spectra of basaltic asteroids (Hardersen et al., 2014, 2015).

Another complication includes the inability of the low-resolution SDSS spectrum to capture the actual shape of an absorption feature in the ~0.9-μm region. As different minerals and mineral mixtures may be spectrally present, the SDSS two-point spectrum at 0.763- and 0.913-μm (the *i'* and *z'* magnitudes) is unable to capture band shape variations within this spectral interval and at wavelengths beyond the *z'* filter magnitude.

Data and Observations. The dataset for this work includes 33 $V_p$ asteroids that were observed between September 2013 and December 2016. Table 1 lists orbital, physical, and albedo information for each asteroid in this work along with the 16 asteroids from Hardersen et al. (2014, 2015). Figure 1 plots the 49 $V_p$ asteroids from this work and from Hardersen et al. (2014, 2015) as a function of orbital inclination vs. semimajor axis (AU). Twenty-five of the $V_p$ asteroids in this work, and 41 $V_p$ asteroids in Table 1, have semi-major axes < 2.5 AU while eight $V_p$ asteroids have semi-major axes > 2.5 AU. These two groups of $V_p$ asteroids are called inner-belt and outer-belt basaltic asteroid candidates, respectively. All the asteroids in Table 1 have NEOWISE-derived effective diameters ($D_{eff}$) <

11 km (Masiero et al., 2011; Mainzer et al., 2016) with a mean effective diameter of 4.872 km. Our target asteroids were also selected based on NEOWISE-derived geometric albedos ($p_v$) that are generally consistent with (4) Vesta ($p_v$ = 0.4228, Tedesco et al., 2002), but span a $p_v$ range from 0.178 to 0.554 with a mean value of 0.343.

All observations and NIR spectra were obtained using the NASA Infrared Telescope Facility (IRTF) on Mauna Kea, Hawai'i, either on-site or remotely. Data were acquired with the SpeX medium-resolution spectrograph and imager using a consistent observational protocol and with consistent instrumental settings (Rayner et al., 2003, 2004). Remote observations with SpeX are identical to on-site observing using a VNC client for remote observers. Observations each night included obtaining ~10-20 spectra per asteroid, 20 spectra of a late F-/G-type main sequence extinction star typically less than 3° away from the asteroid on the sky and spanning an airmass range greater than the asteroid during the observations. Extinction stars are primarily observed to perform spectral telluric corrections during data reduction. About 20-30 spectra of a single solar analog star are also obtained each night, which are used for spectral slope corrections. Flat field images and argon (Ar) spectra are obtained to correct for array pixel sensitivity variations and spectral wavelength calibration. All SpeX data were obtained using the low-resolution ($R \sim 95$) prism mode (0.7 to 2.5 μm) with a 0.8"-wide slit and an open dichroic to obtain the largest spectral range of data possible.

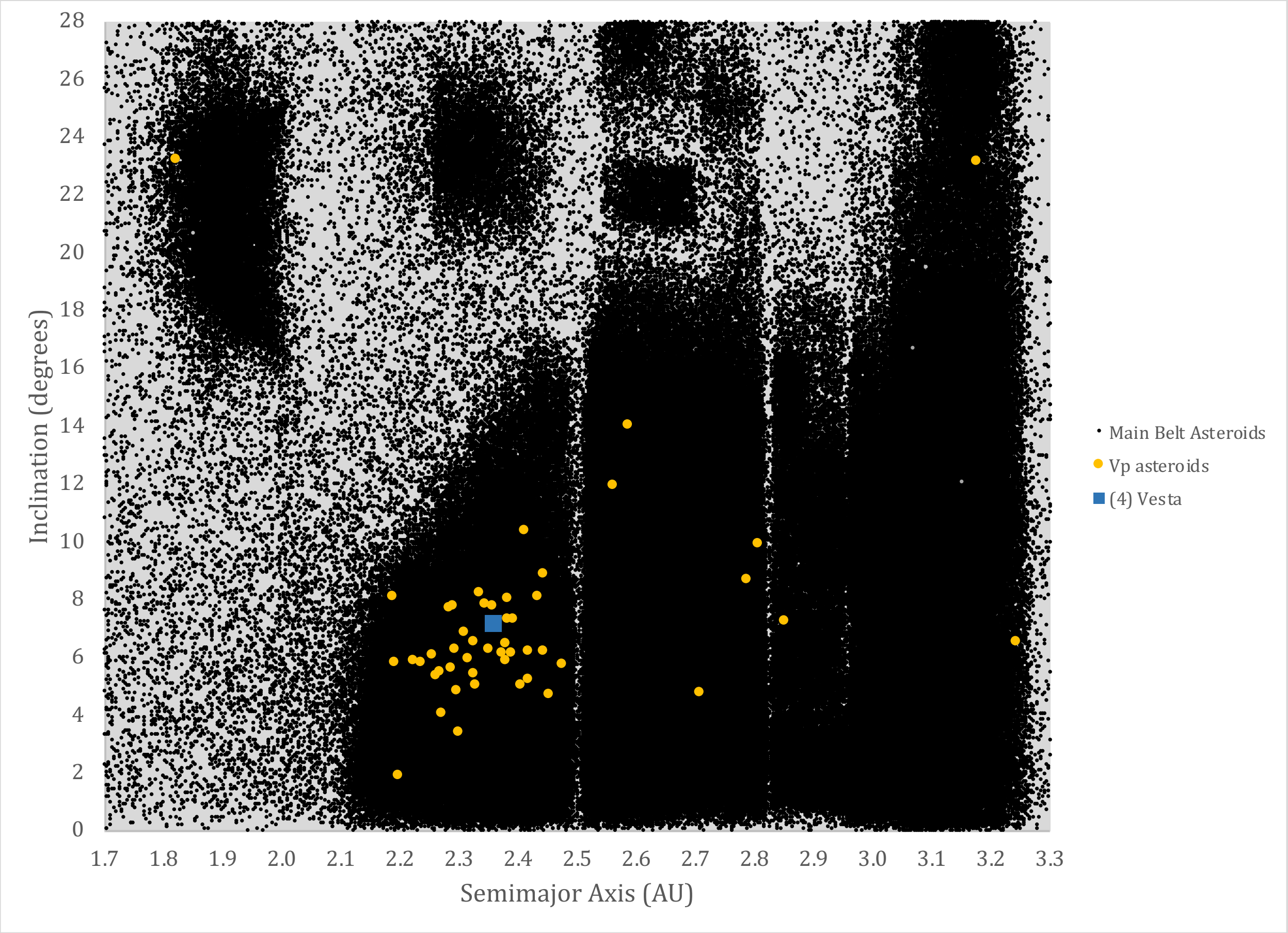

Figure 1. Plot of 49 $V_p$ asteroids from this work and Hardersen et al. (2014, 2015) overlain with the main asteroid belt population, as a function of orbital inclination vs. semimajor axis. Main asteroid belt data obtained from MPCORB data file at the IAU Minor Planet Center.

| Table 1. Orbital elements and physical data for 49 $V_p$ asteroids from this work and Hardersen et al. (2014, 2015). | | | | | | |
|---|---|---|---|---|---|---|
| $D_{eff}$ = *Effective diameter,* $p_v$ = visible albedo; $p_{IR}$ = infrared albedo (Masiero et al., 2011). | | | | | | |
| The first group of asteroids is from this work, the second and third groups from Hardersen et al. (2014, 2015), respectively. | | | | | | |
| | | | | NEOWISE | NEOWISE | NEOWISE |
| Asteroid | $a$ (AU) | $e$ | $i$ (º) | $D_{eff}$ (km) | $p_v$ | $p_{IR}$ |
| (2168) Swope | 2.452 | 0.154 | 4.745 | 8.205±0.058 | 0.263±0.011 | 0.459±0.043 |
| (3715) Stohl | 2.315 | 0.099 | 5.935 | 4.913±0.065 | 0.384±0.068 | 0.509±0.065 |
| (3782) Celle | 2.415 | 0.095 | 5.251 | 5.924±0.23 | 0.503±0.078 | 0.675±0.141 |
| (3849) Incidentia | 2.475 | 0.048 | 5.775 | 5.798±0.125 | 0.398±0.041 | 0.475±0.089 |
| (4055) Magellan | 1.820 | 0.326 | 23.344 | -- | -- | -- |
| (4900) Maymelou | 2.378 | 0.130 | 5.932 | 4.654±0.155 | 0.514±0.046 | 0.772±0.069 |
| (5754) 1992 $FR_2$ | 2.267 | 0.142 | 5.540 | 6.337±0.078 | 0.277±0.031 | 0.567±0.093 |
| (5952) Davemonet | 2.270 | 0.112 | 4.064 | 4.861±0.275 | 0.271±0.055 | 0.812±0.078 |
| (7302) 1993 CQ | 2.807 | 0.181 | 9.954 | 9.56±0.51 | 0.262±0.037 | 0.339±0.038 |
| (8271) Imai | 2.410 | 0.209 | 10.413 | 5.783±0.234 | 0.192±0.028 | 0.288±0.042 |
| (9064) Johndavies | 2.434 | 0.128 | 8.133 | 3.969±0.33 | 0.407±0.18 | 0.611±0.145 |
| (9223) Leifandersson | 2.300 | 0.071 | 3.413 | 4.498±0.176 | 0.381±0.051 | 0.571±0.077 |
| (9368) Esashi | 2.310 | 0.118 | 6.848 | 4.26±0.164 | 0.466±0.043 | 0.699±0.065 |
| (9531) Jean-Luc | 2.235 | 0.187 | 5.816 | 4.176±0.228 | 0.279±0.05 | 0.418±0.076 |
| (10537) 1991 $RY_{16}$ | 2.851 | 0.066 | 7.248 | 7.865±0.269 | 0.313±0.053 | 0.489±0.039 |
| (10666) Feldberg | 2.223 | 0.058 | 5.919 | 3.978±0.038 | 0.233±0.02 | 0.35±0.03 |
| (11341) Babbage | 2.381 | 0.056 | 7.358 | 4.404±0.169 | 0.302±0.044 | 0.452±0.067 |
| (11699) 1998 $FL_{105}$ | 2.406 | 0.083 | 5.078 | 6.447±0.172 | 0.245±0.034 | 0.441±0.045 |
| (12073) Larimer | 2.416 | 0.085 | 6.241 | 2.951±0.227 | 0.456±0.163 | 0.697±0.242 |
| (14390) 1990 $QP_{10}$ | 3.243 | 0.108 | 6.583 | 10.767±0.189 | 0.22±0.057 | 0.33±0.086 |
| (15630) Disanti | 2.326 | 0.115 | 5.461 | 3.548±0.182 | 0.332±0.027 | 0.546±0.056 |
| (16703) 1995 $ER_7$ | 2.441 | 0.172 | 8.933 | 3.132±0.729 | 0.544±0.179 | 0.815±0.268 |
| (17035) Velichko | 2.443 | 0.146 | 6.243 | 4.758±0.314 | 0.283±0.08 | 0.425±0.12 |
| (17480) 1991 $PE_{10}$ | 2.788 | 0.177 | 8.685 | 4.301±0.074 | 0.263±0.065 | 0.394±0.097 |
| (19165) Nariyuki | 2.291 | 0.074 | 7.791 | 3.464±0.196 | 0.487±0.099 | 0.731±0.149 |
| (19738) Callinger | 2.282 | 0.185 | 7.736 | 3.272±0.082 | 0.314±0.056 | 0.472±0.084 |
| (20171) 1996 $WC_2$ | 2.383 | 0.032 | 8.034 | 2.346±0.602 | 0.508±0.286 | 0.763±0.249 |
| (24014) 1999 $RB_{118}$ | 2.560 | 0.190 | 11.967 | 6.513±0.35 | 0.24±0.053 | 0.36±0.079 |
| (25849) 2000 $ET_{107}$ | 2.585 | 0.126 | 14.034 | 5.287±0.171 | 0.276±0.045 | 0.414±0.067 |
| (26417) Michaelgord | 2.256 | 0.058 | 6.082 | 2.325±0.557 | 0.472±0.199 | 0.708±0.237 |
| (27025) 1998 $QY_{77}$ | 2.292 | 0.058 | 6.307 | 2.007±0.599 | 0.439±0.159 | 0.658±0.239 |
| (34698) 2001 $OD_{22}$ | 3.176 | 0.068 | 23.210 | 8.174±0.23 | 0.383±0.066 | 0.567±0.122 |
| (36118) 1999 $RE_{135}$ | 2.709 | 0.035 | 4.797 | 4.998±0.322 | 0.338±0.077 | 0.508±0.116 |
| | | | | | | |
| (3867) Shiretoko | 2.351 | 0.107 | 6.275 | 5.345±0.153 | 0.324±0.058 | 0.487±0.087 |
| (5235) Jean-Loup | 2.298 | 0.141 | 4.848 | 6.709±0.109 | 0.36±0.056 | 0.524±0.164 |
| (5560) Amytis | 2.287 | 0.107 | 5.618 | 4.703±0.041 | 0.29±0.051 | 0.451±0.09 |
| (6331) 1992 $FZ_1$ | 2.359 | 0.134 | 7.763 | 5.321±0.101 | 0.473±0.101 | 0.553±0.115 |
| (6976) Kanatsu | 2.333 | 0.169 | 8.247 | 5.497±0.116 | 0.307±0.042 | 0.46±0.064 |
| (17469) 1991 BT | 2.372 | 0.083 | 6.165 | 5.999±0.173 | 0.258±0.035 | 0.386±0.052 |
| (29796) 1999 $CW_{77}$ | 2.344 | 0.075 | 7.868 | 4.851±0.073 | 0.248±0.062 | 0.392±0.025 |
| (30872) 1992 $EM_{17}$ | 2.328 | 0.113 | 5.060 | 2.995±0.102 | 0.455±0.069 | 0.705±0.108 |
| | | | | | | |
| (2011) Veteraniya | 2.386 | 0.150 | 6.187 | 5.193±0.646 | 0.463±0.1 | 0.695±0.15 |
| (5875) Kuga | 2.379 | 0.049 | 6.469 | 7.465±0.144 | 0.381±0.119 | 0.503±0.084 |
| (8149) Ruff | 2.324 | 0.142 | 6.584 | 4.091±0.132 | 0.554±0.116 | 0.831±0.133 |
| (9147) Kourakuen | 2.191 | 0.106 | 5.816 | 4.854±0.166 | 0.248±0.071 | 0.473±0.084 |
| (9553) Colas | 2.199 | 0.117 | 1.920 | 3.791±0.141 | 0.178±0.029 | 0.266±0.044 |
| (15237) 1988 $RL_6$ | 2.392 | 0.149 | 7.326 | 2.557±0.52 | 0.47±0.254 | 0.704±0.19 |
| (31414) Rotarysusa | 2.260 | 0.153 | 5.376 | 2.547±0.179 | 0.282±0.056 | 0.423±0.084 |
| (32940) 1995 $UW_4$ | 2.189 | 0.136 | 8.111 | 3.351±0.07 | 0.273±0.09 | 0.41±0.135 |

Table 2 describes the respective observational circumstances for the 33 $V_p$ asteroids in this paper. All $V_p$ asteroids were observed with apparent magnitudes within the range from 16.0 to 18.0, cumulative integration times from 400 to 3000 seconds, and total spectra acquired for each asteroid ranging from 4 to 20. Most observing runs experienced photometric, clear weather conditions or with only the minor presence of cirrus clouds and low relative humidity. Exceptions include the observing run of June 16, 2015 UT, which experienced cirrus across the sky during the entire night and moderate relative humidity of ~54% late in the observing run; the observing run of June 17, 2015 UT, which experienced moderate-to-high values of relative humidity from 48% to 74% and cirrus throughout the sky at the beginning of the night; and December 11, 2015 UT that experienced poor atmospheric seeing (~1.5" to 2.0").

Data Reduction. Production of average, normalized NIR reflectance spectra occurs using a combination of IDL-based Spextool software (Cushing et al., 2004) and Microsoft Excel. Spextool is specifically designed for IRTF SpeX data and is utilized in the first phase of the data reduction process. Tasks that are accomplished with Spextool include removal of the background sky signal, telluric corrections, channel shifting, flat field and wavelength calibrations, and spectral averaging routines. Successful background sky subtraction is linked to an observational approach that obtains spectra in an *ABBAABBAAB…* sequence at the two slit positions, *A* and *B*, which allows *AB* spectral pairs to be subtracted to efficiently remove the background sky flux contribution. Channel shifting is also important to ensure that all spectra properly overlay and cover the same pixels in each spectrum. Not performing this function introduces unwanted noise into the resulting average spectrum.

Telluric corrections are the most important part of the data reduction because correcting for the atmospheric water vapor absorptions at ~1.4- and ~1.9-μm is vital for producing high-quality, interpretable average spectra. This task is also linked to a specific observational strategy where the extinction star for a given asteroid is observed prior to, and after, the asteroid observations. This sequence is performed to ensure that the airmass range of the star exceeds and encompasses the airmass range of the asteroid, which allows effective empirical modeling of the night sky and removal of the telluric contribution.

Table 2. Observational circumstances for 33 $V_p$-type asteroids in this work.

| Asteroid | Observation Date (UT) | Apparent *V* Magnitude | (º) Phase Angle | Total Integration Time (sec) | Total Spectra | Extinction Star | Solar Analog Star |
|---|---|---|---|---|---|---|---|
| (2168) Swope | 1/16/15 | 17.49 | 16.4 | 1600 | 8 | BD+02 2405 | SAO 120107 |
| (2168) Swope | 1/17/15 | 17.47 | 16.2 | 1400 | 7 | BD+02 2405 | SAO 120107 |
| (2168) Swope | 1/19/15 | 17.43 | 15.7 | 2400 | 12 | BD+02 2405 | SAO 120107 |
| (3715) Stohl | 1/15/15 | 16.19 | 3.4 | 2000 | 10 | HD 65523 | SAO 120107 |
| (3715) Stohl | 1/19/15 | 16.10 | 2.0 | 1600 | 8 | HD 65523 | SAO 120107 |
| (3782) Celle | 1/15/15 | 17.16 | 25.2 | 800 | 4 | HD 116667 | SAO 120107 |
| (3782) Celle | 6/16/15 | 16.45 | 24.7 | 2000 | 10 | SAO 157732 | SAO 120107 |
| (3782) Celle | 6/22/15 | 16.55 | 25.7 | 2000 | 10 | SAO 157732 | SAO 120107 |
| (3849) Incidentia | 1/16/15 | 17.15 | 21.8 | 1600 | 8 | HD 104800 | SAO 120107 |
| (3849) Incidentia | 1/17/15 | 17.13 | 21.6 | 1800 | 9 | HD 104800 | SAO 120107 |
| (4055) Magellan | 6/16/15 | 16.02 | 47.9 | 2000 | 10 | SAO 69926 | SAO 120107 |
| (4900) Maymelou | 1/26/16 | 17.46 | 8.9 | 2000 | 10 | HD 50060 | SAO 93936 |
| (5754) 1992 FR$_2$ | 1/19/15 | 16.47 | 11.5 | 1000 | 5 | HIP 28216 | SAO 120107 |
| (5952) Davemonet | 6/16/15 | 16.74 | 17.9 | 3000 | 15 | SAO 189480 | SAO 120107 |
| (7302) 1993 CQ | 1/26/16 | 16.48 | 19.2 | 2000 | 10 | HD 30047 | SAO 93936 |
| (8271) Imai | 6/16/15 | 16.14 | 7.3 | 2000 | 10 | SAO 141903 | SAO 120107 |
| (8271) Imai | 6/17/15 | 16.13 | 7.2 | 2000 | 10 | SAO 141903 | SAO 120107 |
| (9064) Johndavies | 1/26/16 | 17.15 | 10.2 | 2400 | 12 | HD 87776 | SAO 93936 |
| (9223) Leifandersson | 6/16/15 | 16.18 | 8.8 | 2000 | 10 | SAO 162034 | SAO 120107 |
| (9223) Leifandersson | 6/17/15 | 16.15 | 8.3 | 2000 | 10 | SAO 162034 | SAO 120107 |
| (9368) Esashi | 6/17/15 | 17.12 | 11.4 | 2000 | 10 | HD 186163 | SAO 120107 |
| (9531) Jean-Luc | 1/26/16 | 17.09 | 6.3 | 2800 | 14 | HD 62043 | SAO 93936 |
| (10537) 1991 RY$_{16}$ | 12/29/16 | 17.04 | 7.7 | 1600 | 8 | SAO 76855 | SAO 93936 |
| (10666) Feldberg | 1/16/15 | 16.14 | 4.5 | 1600 | 8 | HD 68017 | SAO 120107 |
| (10666) Feldberg | 1/19/15 | 16.14 | 4.5 | 1400 | 7 | HD 68017 | SAO 120107 |
| (11341) Babbage | 1/26/16 | 16.70 | 13.5 | 2400 | 12 | HD 93653 | SAO 93936 |
| (11699) 1998 FL$_{105}$ | 9/4/14 | 16.40 | 5.1 | 1200 | 10 | SAO 128229 | HD 28099 |
| (12073) Larimer | 9/2/14 | 17.04 | 5.5 | 480 | 4 | SAO 128164 | HD 28099 |
| (14390) 1990 QP$_{10}$ | 9/13/13 | 17.17 | 15.0 | 2400 | 20 | SAO 189379 | SAO 93936 |
| (15630) Disanti | 12/29/16 | 16.45 | 5.7 | 2000 | 10 | SAO 77406 | SAO 93936 |
| (16703) 1995 ER$_7$ | 6/16/15 | 16.91 | 18.9 | 1400 | 7 | SAO 144197 | SAO 120107 |
| (17035) Velichko | 6/22/15 | 17.08 | 11.8 | 1600 | 8 | SAO 159838 | SAO 120107 |
| (17480) 1991 PE$_{10}$ | 1/2/16 | 17.58 | 7.3 | 2600 | 13 | SAO 58270 | -- |
| (19165) Nariyuki | 1/19/15 | 16.71 | 14.8 | 2000 | 10 | BD+36 1276 | SAO 120107 |
| (19738) Callinger | 6/17/15 | 17.12 | 14.8 | 2400 | 12 | SAO 188854 | SAO 120107 |
| (20171) 1996 WC$_2$ | 1/26/16 | 17.67 | 6.9 | 1600 | 8 | HD 82069 | SAO 93936 |
| (24014) 1999 RB$_{118}$ | 6/17/15 | 17.35 | 8.2 | 2000 | 10 | SAO 188103 | SAO 120107 |
| (25849) 2000 ET$_{107}$ | 1/2/16 | 17.48 | 13.7 | 2000 | 10 | HD 27947 | -- |
| (26417) Michaelgord | 12/29/16 | 17.59 | 2.8 | 2800 | 14 | SAO 78788 | SAO 93936 |
| (27025) 1998 QY$_{77}$ | 12/21/06 | 17.36 | 4.5 | 2800 | 14 | HD 261730 | SAO 93936 |
| (34698) 2001 OD$_{22}$ | 1/15/15 | 17.34 | 10.3 | 2000 | 10 | TYC 4894-595-1 | SAO 120107 |
| (34698) 2001 OD$_{22}$ | 1/16/15 | 17.32 | 10.0 | 1200 | 6 | TYC 4894-595-1 | SAO 120107 |
| (36118) 1999 RE$_{135}$ | 1/15/15 | 17.43 | 9.9 | 1200 | 6 | HD 82906B | SAO 120107 |
| (36118) 1999 RE$_{135}$ | 1/19/15 | 17.33 | 8.3 | 2000 | 10 | HD 82906B | SAO 120107 |

Once background subtraction, telluric corrections, and channel shifting have been successfully accomplished, an average of the asteroid spectra relative to the extinction star is performed. The same process is then performed for the solar analog star. These two average spectra are then converted to text files and exported into Microsoft Excel. For the solar analog star spectrum relative to the extinction star, any remnant telluric features are removed and replaced by a smoothed spectrum. This is a standard procedure and is justified based on the lack of broad absorption features present in stellar spectra. The resulting average unnormalized asteroid spectrum is described in the following qualitative equation:

*Average Asteroid/Solar Analog = (Asteroid/Ext. Star) / (Solar Analog/Ext. Star)*

*Average Asteroid/Solar Analog* is the final unnormalized average asteroid spectrum, *(Asteroid/Ext. Star)* is the average asteroid spectrum relative to the extinction star, and *(Solar Analog/Ext. Star)* is the average solar analog star spectrum relative to the extinction star. The average asteroid reflectance spectrum is then normalized to a flux of 1.0 at 1.5 μm, which is a wavelength position along the overall continuum and away from any spectral absorption features.

Data Analysis. The results of the data reduction effort are shown in Figure 2, which displays 44 average NIR spectra for the 33 $V_p$ asteroids in this work. The asteroids are listed numerically based on their number designation and by the UT date of observation. Each asteroid has at least one average spectrum in Figure 2, but several asteroids were observed on multiple nights and the spectra from all nights are shown for comparison.

Determining whether a $V_p$ classified asteroid is likely to have a basaltic nature is evaluated initially on a visual examination of each asteroid's reflectance spectrum and later, as appropriate, by more rigorous spectral band parameter analysis. Asteroids with an average reflectance spectrum that do not display spectral absorption features are identified as not being basaltic due to the lack of the two prominent and deep pyroxene spectral absorption bands at ~0.9- and ~1.9-μm present in basaltic achondrite NIR spectra and the NIR spectrum of (4) Vesta (Gaffey, 1997; Reddy, 2011). Those asteroids that display spectral absorption features, but absorptions that are significantly different from pyroxene absorptions (i.e., weaker absorptions, differences in spectral band shapes, etc.) are flagged as potentially non-basaltic, but are subjected to band parameter analysis. Finally, those asteroids displaying the canonical pyroxene absorptions are subjected to band parameter analysis. For the latter two cases, the three-tiered test is applied to determine the basaltic affinity for each asteroid.

Spectral band parameter analysis is a technique to extract, measure, and quantify the different absorption features of a NIR spectrum. This technique can potentially be applied to all asteroid NIR spectra with measurable spectral absorption features but is most useful for well-calibrated pyroxene-dominated, olivine-dominated, or pyroxene-olivine dominated spectra (Burns, 1993; Gaffey et al., 1993; Sanchez et al., 2014; Hardersen et al., 2014, 2015). Laboratory pyroxene and basaltic achondrite calibrations are the most relevant calibrations for this work because pyroxene is the dominant mafic silicate component in basaltic achondrites and the Howardite-Eucrite-Diogenite (HED) clan (Mittlefehldt et al., 1998).

It is also important to note that Type B pyroxenes (orthopyroxenes, low-Ca clinopyroxenes) are the relevant types of pyroxenes for basaltic achondrites and display both primary absorption features as compared to the high-Ca Type A pyroxenes (generally along the diopside-hedenbergite trend in the pyroxene quadrilateral) that often only include the ~0.9-μm absorption (Adams, 1974; Schade et al., 2004). A mineral chemistry calibration does not exist for high-Ca Type A clinopyroxenes, although Schade et al. (2004) suggests a possible association may exist between the 1.15-μm feature in the pyroxene M1 coordination site and Fe content.

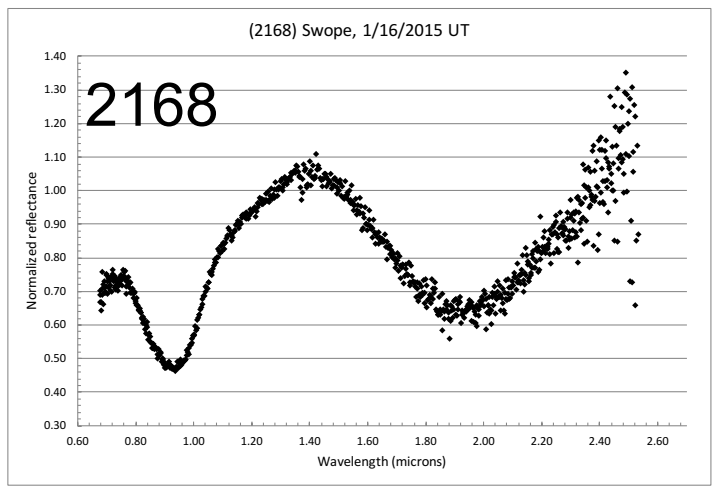
(2168) Swope, 1/16/2015 UT
2168
Normalized reflectance
Wavelength (microns)

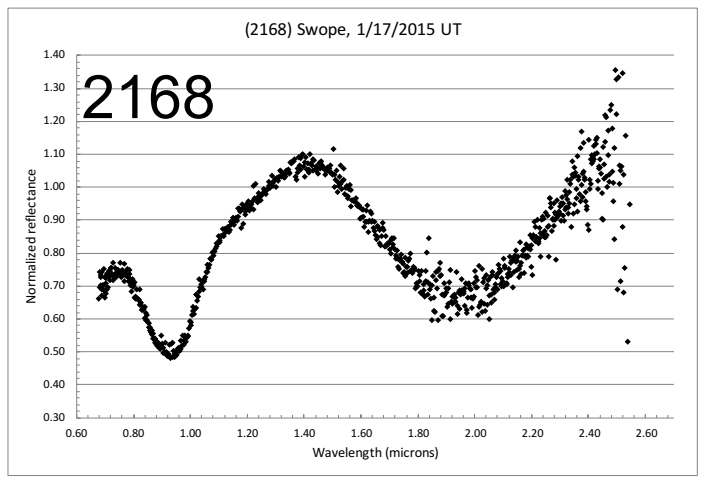
(2168) Swope, 1/17/2015 UT
2168
Normalized reflectance
Wavelength (microns)

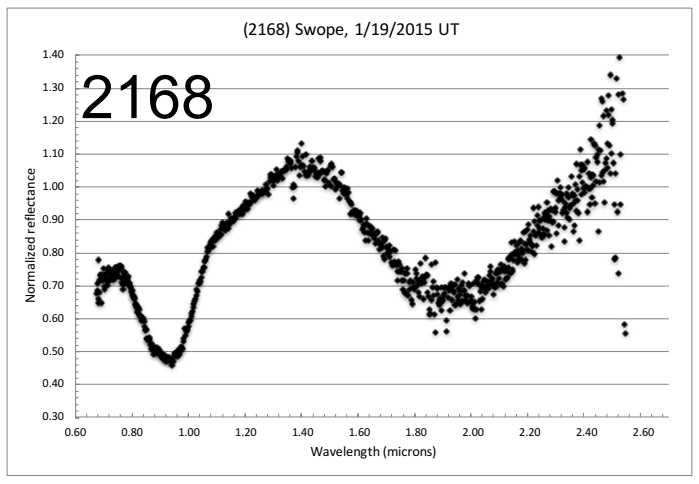
(2168) Swope, 1/19/2015 UT
2168
Normalized reflectance
Wavelength (microns)

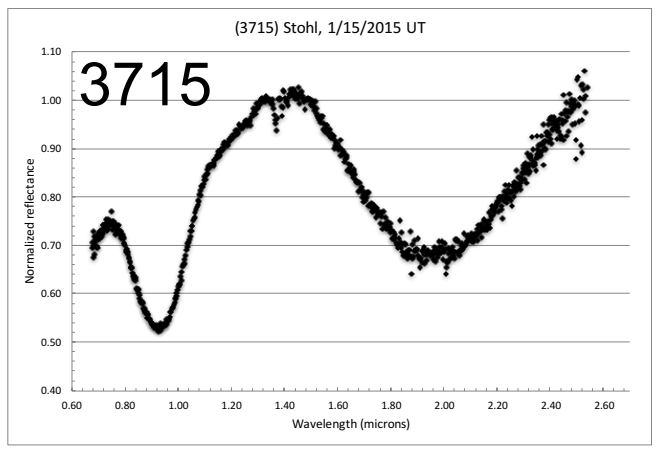
(3715) Stohl, 1/15/2015 UT
3715
Normalized reflectance
Wavelength (microns)

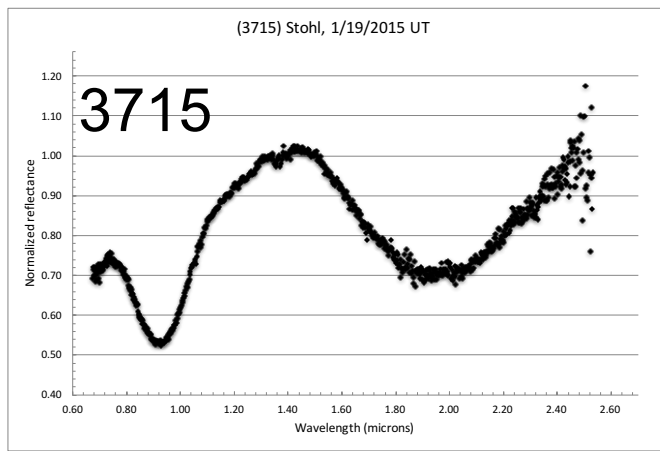
(3715) Stohl, 1/19/2015 UT
3715
Normalized reflectance
Wavelength (microns)

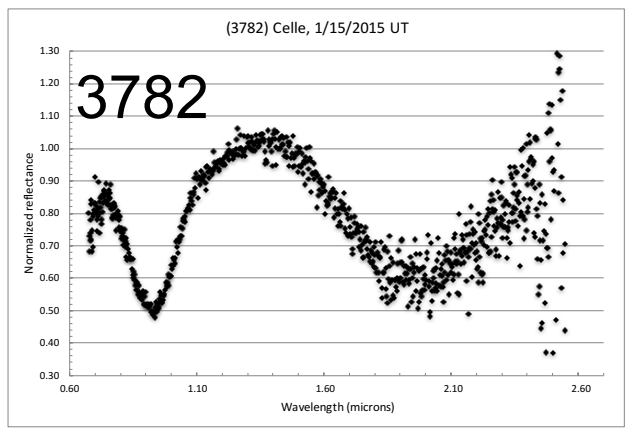
(3782) Celle, 1/15/2015 UT
3782
Normalized reflectance
Wavelength (microns)

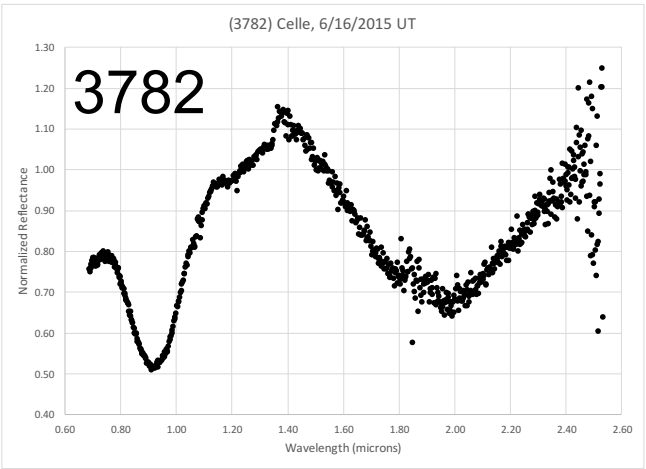
(3782) Celle, 6/16/2015 UT
3782
Normalized Reflectance
Wavelength (microns)

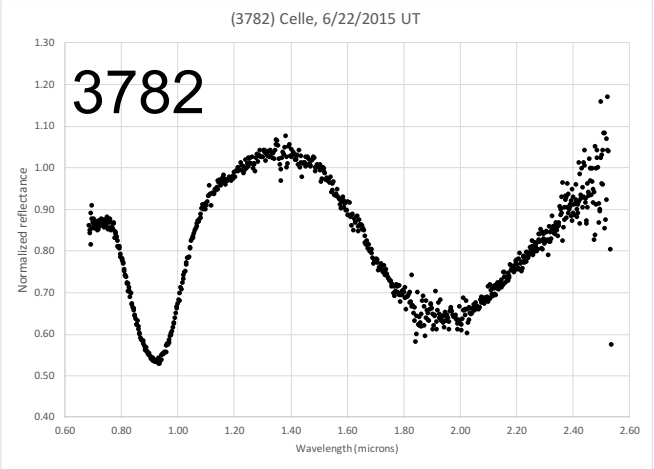
(3782) Celle, 6/22/2015 UT
3782
Normalized reflectance
Wavelength (microns)

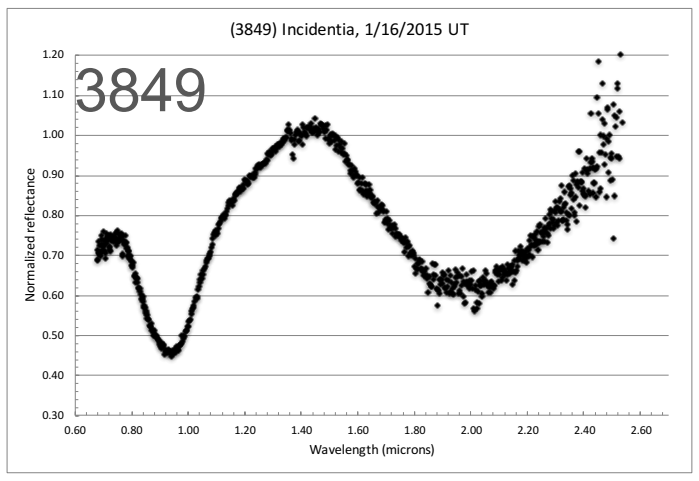
(3849) Incidentia, 1/16/2015 UT
3849
Normalized reflectance
Wavelength (microns)

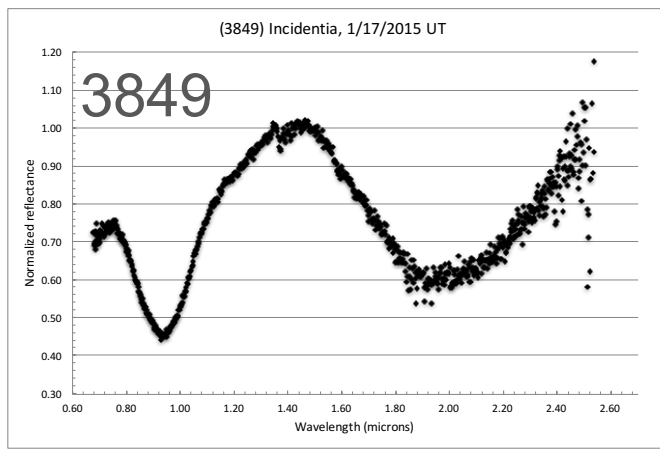
(3849) Incidentia, 1/17/2015 UT
3849
Normalized reflectance
Wavelength (microns)

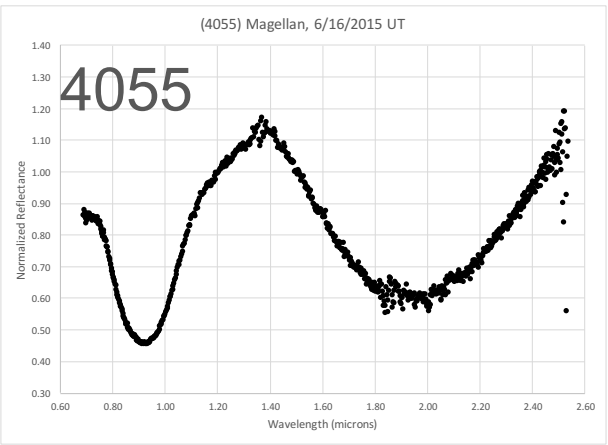
(4055) Magellan, 6/16/2015 UT
4055
Normalized Reflectance
Wavelength (microns)

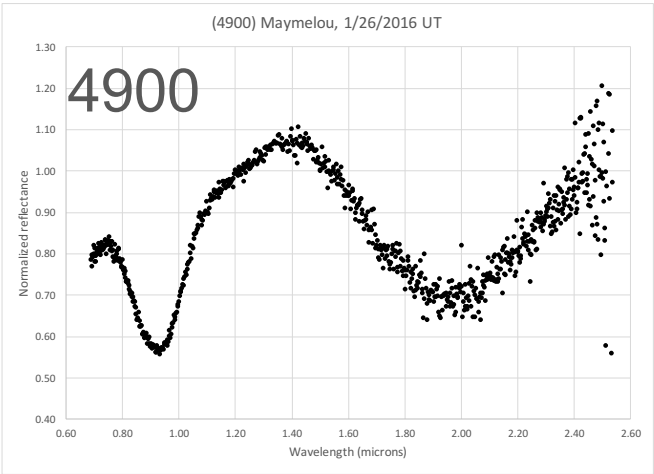
(4900) Maymelou, 1/26/2016 UT
4900
Normalized reflectance
Wavelength (microns)

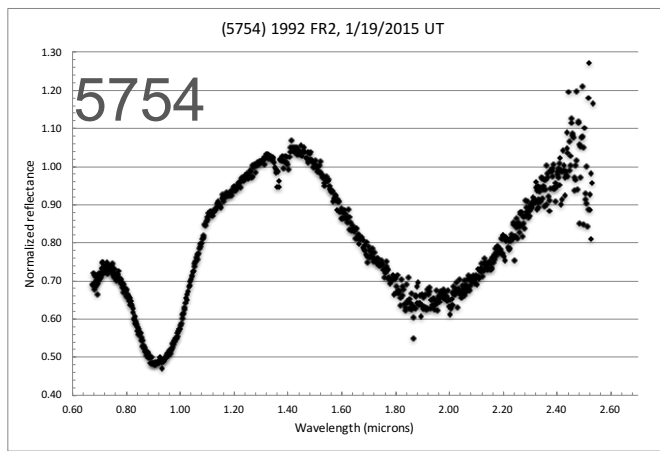
(5754) 1992 FR2, 1/19/2015 UT
5754
Normalized reflectance
Wavelength (microns)

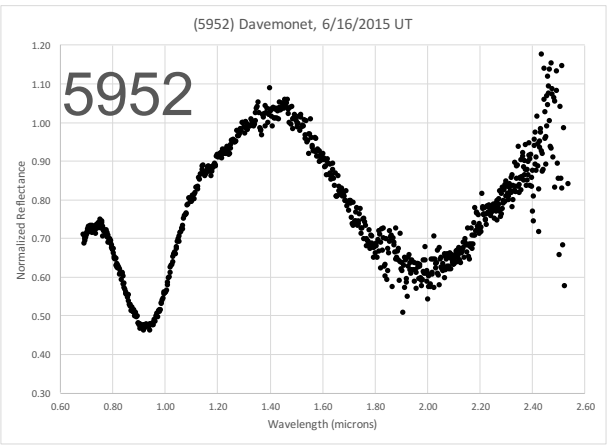
(5952) Davemonet, 6/16/2015 UT
5952
Normalized Reflectance
Wavelength (microns)

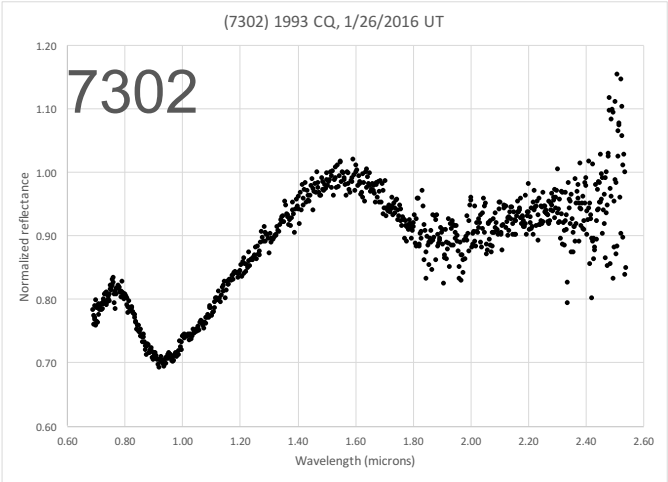
(7302) 1993 CQ, 1/26/2016 UT
7302
Normalized reflectance
Wavelength (microns)

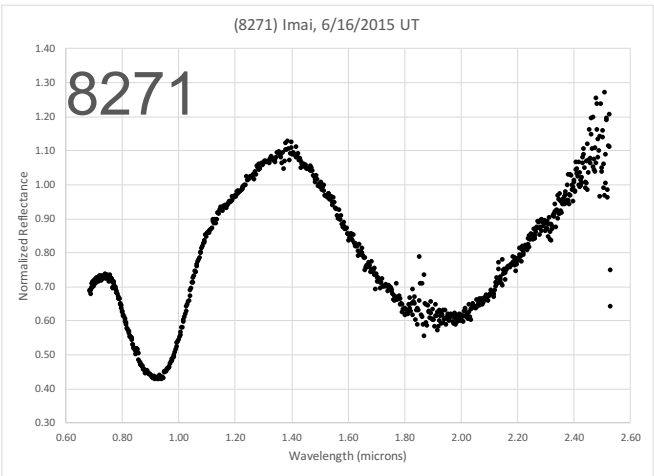
(8271) Imai, 6/16/2015 UT
8271
Normalized Reflectance
Wavelength (microns)

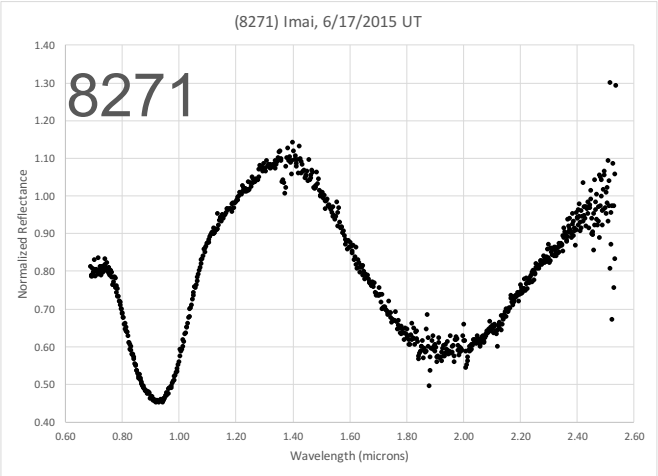
(8271) Imai, 6/17/2015 UT
8271
Normalized Reflectance
Wavelength (microns)

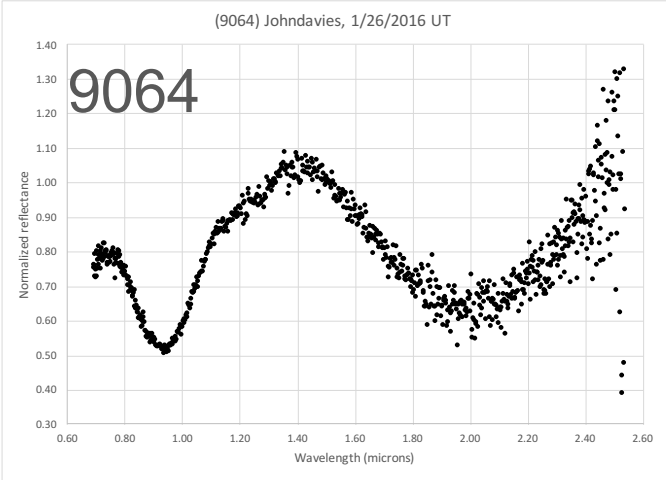
(9064) Johndavies, 1/26/2016 UT
9064
Normalized reflectance
Wavelength (microns)

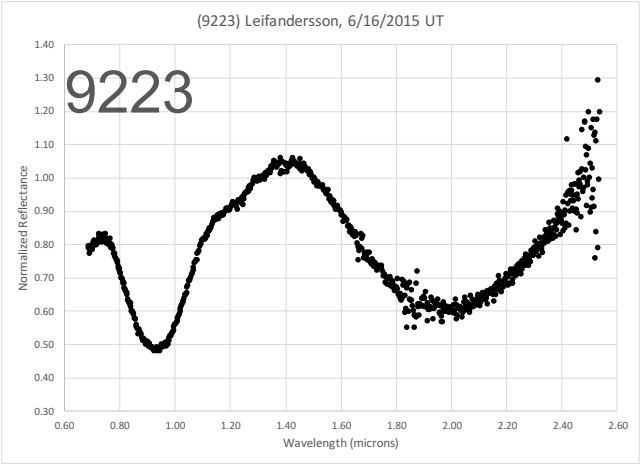
(9223) Leifandersson, 6/16/2015 UT
9223
Normalized Reflectance
Wavelength (microns)

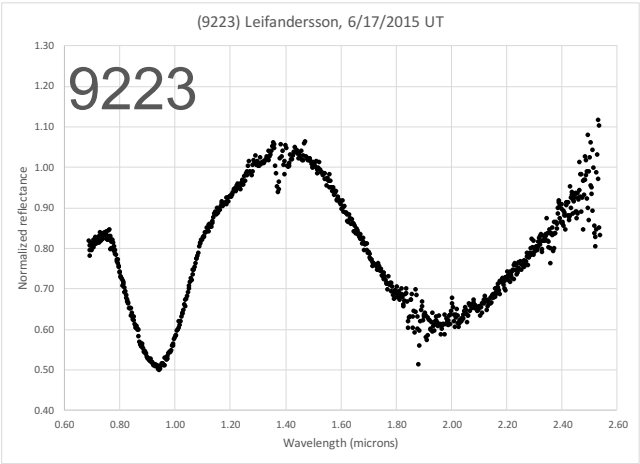
(9223) Leifandersson, 6/17/2015 UT
9223
Normalized reflectance
Wavelength (microns)

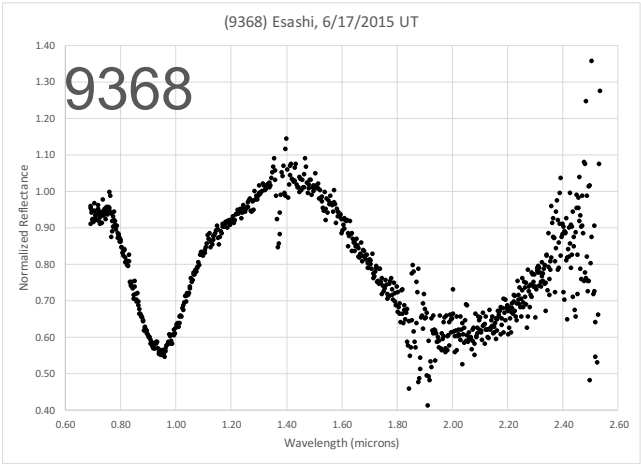
(9368) Esashi, 6/17/2015 UT
9368
Normalized Reflectance
Wavelength (microns)

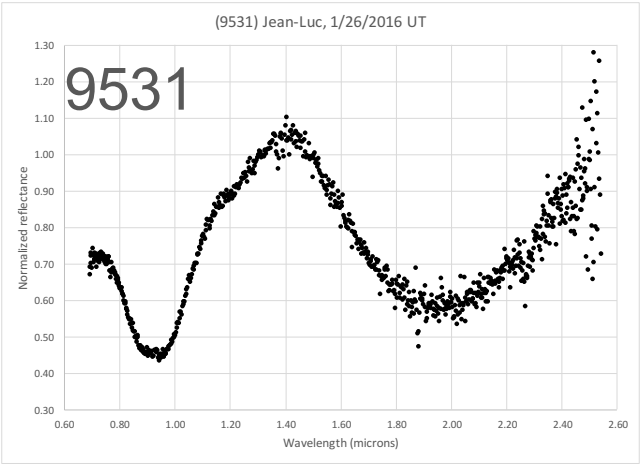
(9531) Jean-Luc, 1/26/2016 UT
9531
Normalized reflectance
Wavelength (microns)

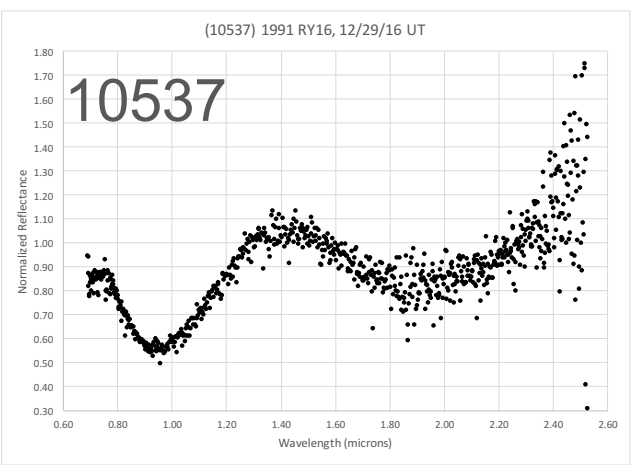
(10537) 1991 RY16, 12/29/16 UT
10537
Normalized Reflectance
Wavelength (microns)

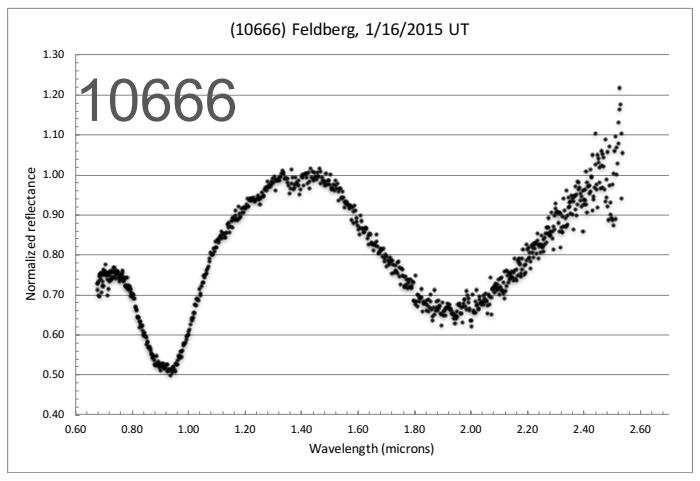
(10666) Feldberg, 1/16/2015 UT
10666
Normalized reflectance
Wavelength (microns)

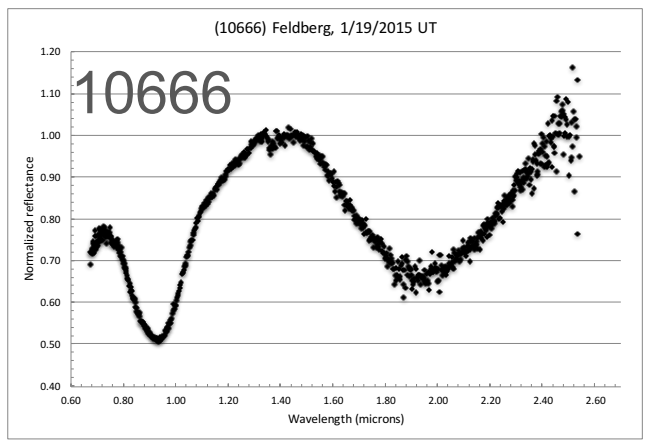
(10666) Feldberg, 1/19/2015 UT
10666
Normalized reflectance
Wavelength (microns)

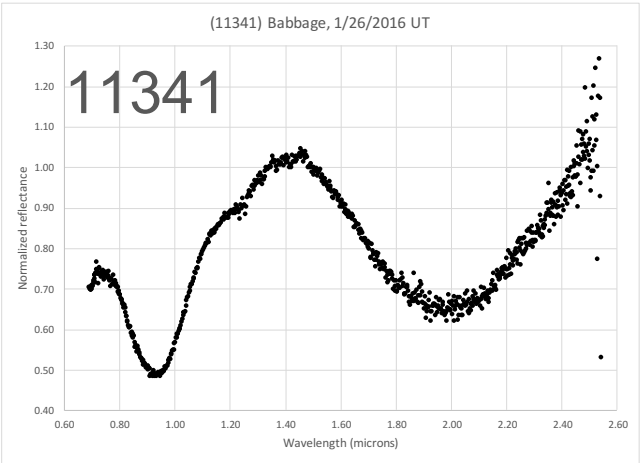
(11341) Babbage, 1/26/2016 UT
11341
Normalized reflectance
Wavelength (microns)

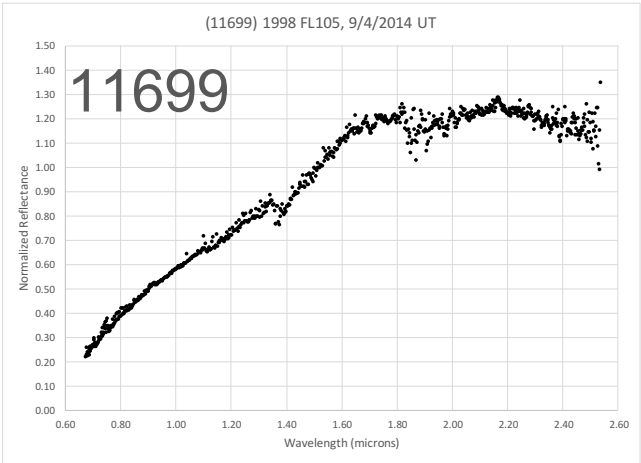
(11699) 1998 FL105, 9/4/2014 UT
11699
Normalized Reflectance
Wavelength (microns)

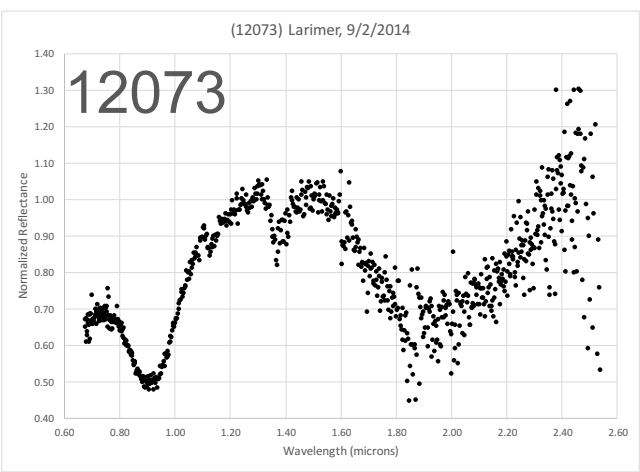
(12073) Larimer, 9/2/2014
12073
Normalized Reflectance
Wavelength (microns)

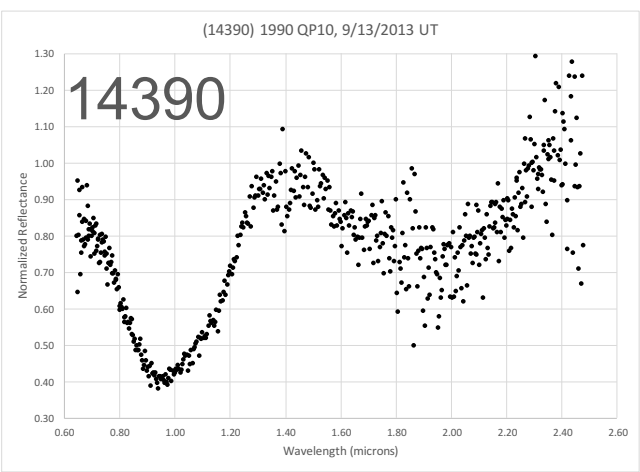
(14390) 1990 QP10, 9/13/2013 UT
14390
Normalized Reflectance
Wavelength (microns)

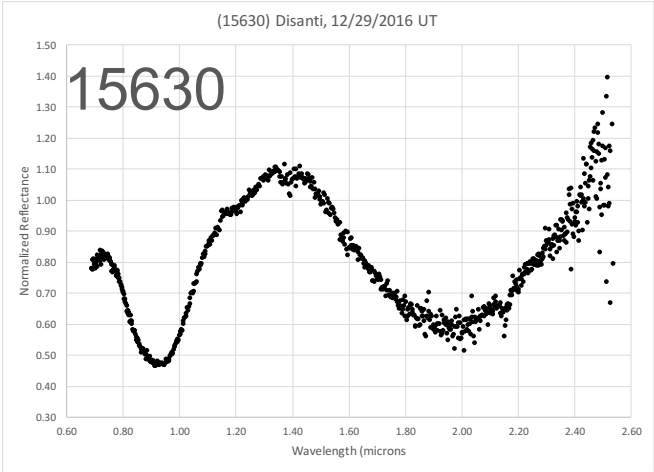
(15630) Disanti, 12/29/2016 UT
15630
Normalized Reflectance
Wavelength (microns)

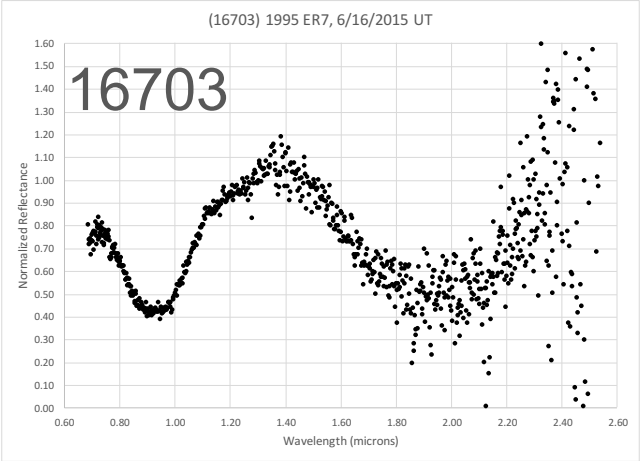
(16703) 1995 ER7, 6/16/2015 UT
16703
Normalized Reflectance
Wavelength (microns)

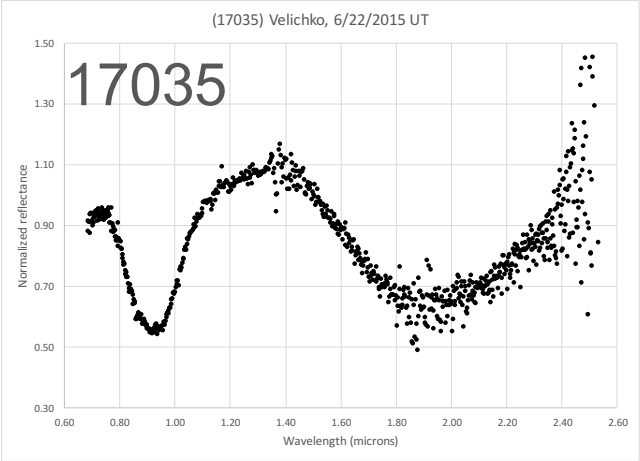
(17035) Velichko, 6/22/2015 UT
17035
Normalized Reflectance
Wavelength (microns)

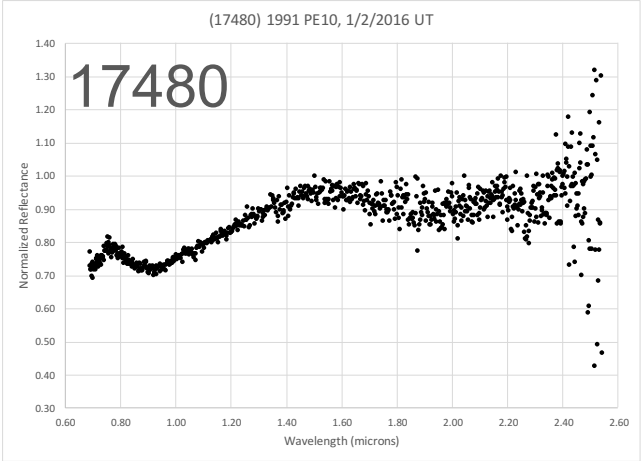
(17480) 1991 PE10, 1/2/2016 UT
17480
Normalized Reflectance
Wavelength (microns)

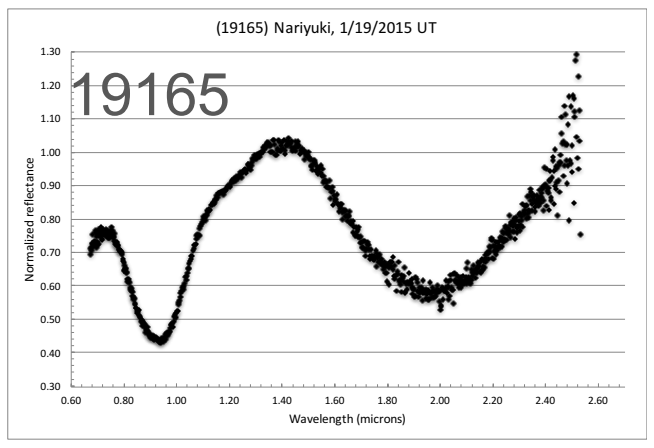
(19165) Nariyuki, 1/19/2015 UT
19165
Normalized reflectance
Wavelength (microns)

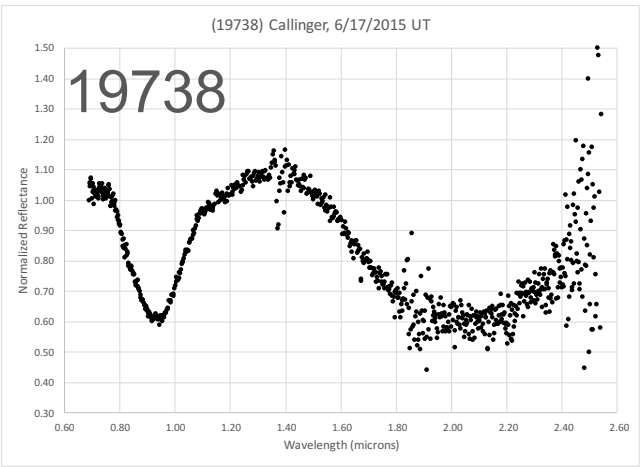
(19738) Callinger, 6/17/2015 UT
19738
Normalized Reflectance
Wavelength (microns)

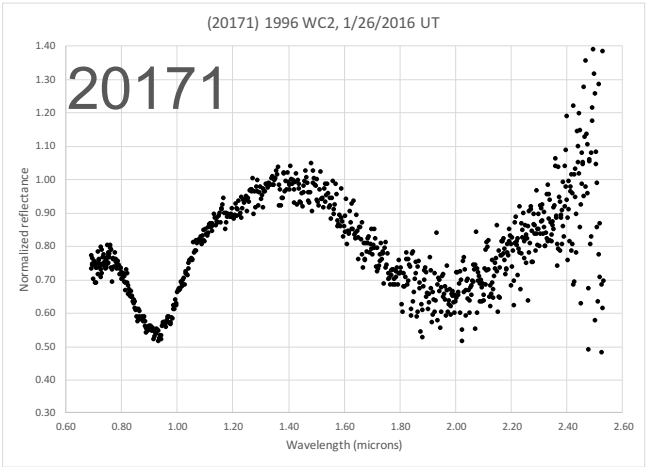
(20171) 1996 WC2, 1/26/2016 UT
20171
Normalized Reflectance
Wavelength (microns)

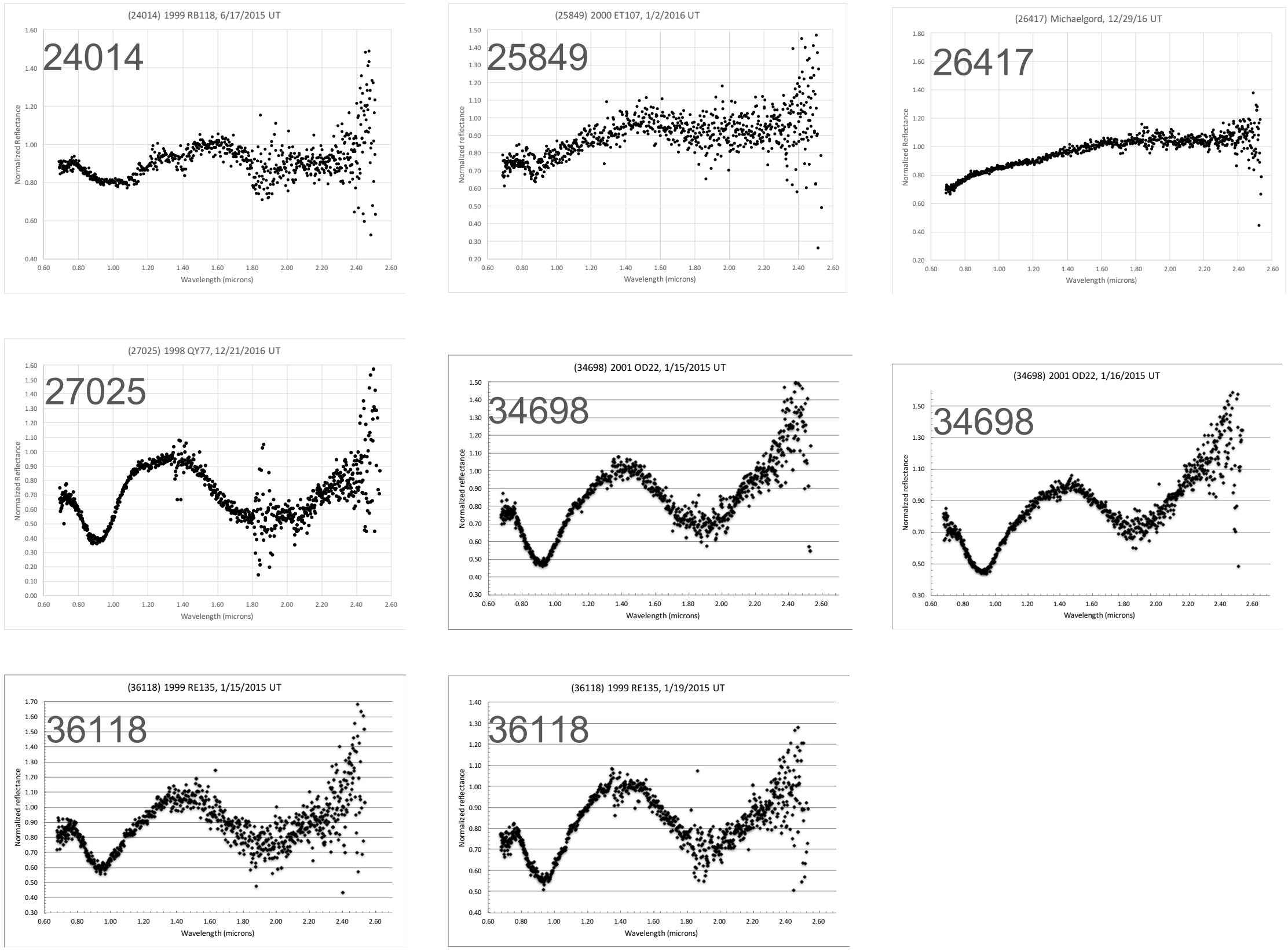

Figure 2. Average NIR reflectance spectra for 33 $V_p$ asteroids in this work.

The primary spectral band parameters that are quantified include absorption band centers, band areas, and Band Area Ratios (BAR). An absorption band center is the minimum reflectance of a continuum-removed absorption feature. This is produced by creating a linear continuum across the absorption meeting at the local maximum on each side of the absorption. A ratio of the linear continuum to the absorption **feature** produces the continuum-removed feature, **which removes the effect of any positive or negative slope in this region of the NIR spectrum. The continuum-removed absorption** can then be directly linked to pyroxene mineral chemistries (Burns, 1993). Analysis of Type B pyroxenes with a range of mineral chemistries has shown that band centers systematically shift to longer wavelengths with higher Fe and Ca content (Adams, 1974; Cloutis et al., 1986; Burns, 1993).

Band areas are then calculated from the areas of individual, continuum-removed absorptions. BAR is the ratio of the areas of the ~1.9-μm (i.e., Band II) feature to the ~0.9-μm (i.e., Band I) feature. A Band I vs. BAR calibration exists for olivine-orthopyroxene mixtures (Cloutis et al., 1986; Gaffey et al., 2002) where larger BAR values indicate a greater relative pyroxene abundance compared to olivine with basaltic achondrites typically exhibiting BAR values > 1.5 (Cloutis et al., 1986; Gaffey et al., 2002). Absorption band depth is another commonly measured spectral band parameter, which represents the percentage decrease in reflectance of an absorption relative to the isolated continuum (e.g., Band Depth (%) = 1 – $X$ = $Y$, where $X$ is the relative reflectance value of the minimum reflectance of the absorption band and $Y$ is the continuum-removed absorption band depth represented as a percentage). Band depths are not diagnostic of mineralogy and can vary based on the asteroid's phase angle during observations, space weathering on an asteroid's surface, or a surface metal component (Britt and Pieters, 1988; Blewett et al., 2015; Cloutis et al., 1990, 2015; Sanchez et al., 2014).

Our band parameter analysis utilizes a set of MATLAB routines (Reddy et al., 2012a) and the IDL-based Spectral Analysis Routine for Asteroids (SARA) (Lindsay et al., 2015). While both sets of code extract the same band parameters, the MATLAB method is semi-automated and the SARA method is completely automated. The only difference in the methods is that we used the SARA routines with the default Band II long wavelength limit of 2.50-μm, while we manually chose the Band II long-wavelength limit in MATLAB for each asteroid based on the furthest extent of the measurable absorption.

Table 3 displays the MATLAB, SARA, and average MATLAB/SARA-derived and temperature-corrected band centers. **To determine band centers, polynomial fits with different orders were used to measure the lower region of each absorption feature. The semi-automated MATLAB routines (Hardersen et al., 2014, 2015, and references therein) utilize 3$^{rd}$ and 4$^{th}$ order polynomials to obtain at least 20 measurements of the lower third of each absorption feature. The results for both orders were consistent across our set of measurements. The automated SARA routines produce 3$^{rd}$, 4$^{th}$, and 5$^{th}$ order polynomial fits, but we only used the 4$^{th}$ and 5$^{th}$ order fits as the 3$^{rd}$ order fits often fit the data poorly and were not consistent with the higher-order fits. Multiple runs of each NIR spectrum through SARA were also conducted to ensure consistency of derived band centers and BARs. The results in Table 3 show that the Band I averages derived from the MATLAB and SARA routines are very consistent. The results for the Band II averages show larger deviations, which is primarily attributable to greater scatter in the data in this spectral region due to inherently lower Signal-To-Noise (SNR) at longer wavelengths in the NIR spectra.**

**The HED band center data in Figure 3 and Figure 4 were measured using the same approach as in this work (and Hardersen et al., 2014, 2015) while also using very similar routines. Reddy et al. (2012b) used a Python code to measure the HED band center data that is very similar to the MATLAB code used by this work, but is more automated than the MATLAB code.**

Temperature corrections from (Reddy et al., 2012b) are applied to both the Band I and Band II centers with the largest corrections applied to Band II centers. MATLAB band centers are averages of ~40 measurements and SARA results are averages of three runs to ensure repeatability and consistency of the results. Band center errors encompass the largest deviations that exist among the MATLAB and SARA derived values and exceed the 1σ standard deviations of both the individual average MATLAB and SARA results.

Table 4 displays the resulting average asteroid surface pyroxene chemistries using the Gaffey et al. (2002), Burbine et al. (2009), and the average chemistries derived from both calibrations. Table 4 also includes the average BAR values obtained from an average of the MATLAB and SARA results.

Interpretations and Results. Of the 33 $V_p$ asteroids in this work, 25 asteroids reside in the inner main-belt with $a$ < 2.5 AU while eight asteroids have $a$ > 2.5 AU. Two of the inner-belt $V_p$ asteroids, (11699) 1998 $FL_{105}$ and (26417) Michaelgord are **potentially** not basaltic asteroids and do not have analogs with the HED meteorites. The remaining 23 inner-belt $V_p$ asteroids are consistent with a basaltic composition and have likely analogs with the HED meteorites.

The eight outer-belt $V_p$ asteroids include two asteroids, (34698) 2001 $OD_{22}$ and (36118) 1999 $RE_{135}$, that have a likely basaltic composition. The other six $V_p$ asteroids are not consistent with a basaltic composition and display significant spectral variability and likely surface compositional and meteorite analog diversity.

Inner-belt $V_p$ Asteroids. The group of inner-belt $V_p$ asteroids in this work are mostly consistent with a basaltic composition and associations with specific HED meteorite analogs. We distinguish three sub-groups among these inner-belt $V_p$ asteroids: 1) those asteroids that display uniformly consistent spectral and mineralogical evidence in support of a basaltic surface composition and association with an HED meteorite type, 2) those asteroids also exhibiting consistent evidence for a basaltic composition and HED meteorite analogs with the exception of a difference in Band I vs. Band II positions in Figure 3A/3B, and 3) those asteroids that are spectrally and mineralogically inconsistent with a basaltic composition. The second group of asteroids is consistent with other $V_p$ asteroids from Hardersen et al. (2014, 2015) that also plot just off the Band I vs. Band II plot, but still exhibit an HED meteorite analog.

| Table 3. Absorption band centers for the $V_p$ asteroids in this work that exhibit absorptions across the 0.7 to 2.5 μm spectral range. Band centers are temperature-corrected and display the MATLAB, SARA, and average results. MATLAB results for NIR spectra of (3782) Celle and (10537) 1991 RY16 from Moskovitz et al. (2010) are shown for comparison. | | | | | | |
|---|---|---|---|---|---|---|
| | MATLAB (microns) | MATLAB (microns) | SARA (microns) | SARA (microns) | Average (microns) | Average (microns) |
| Asteroid | Band I center | Band II center | Band I center | Band II center | Band I center | Band II center |
| (2168) Swope | 0.947 ± 0.003 | 1.953 ± 0.006 | 0.943 | 1.960 | 0.945 ± 0.003 | 1.957 ± 0.006 |
| (2168) Swope | 0.948 ± 0.003 | 1.973 ± 0.009 | 0.943 | 1.976 | 0.946 ± 0.003 | 1.973 ± 0.009 |
| (2168) Swope | 0.942 ± 0.003 | 1.953 ± 0.010 | 0.939 | 1.957 | 0.941 ± 0.003 | 1.955 ± 0.010 |
| (3715) Stohl | 0.943 ± 0.003 | 1.973 ± 0.004 | 0.942 | 1.982 | 0.943 ± 0.003 | 1.978 ± 0.005 |
| (3715) Stohl | 0.940 ± 0.003 | 1.973 ± 0.008 | 0.931 | 1.978 | 0.940 ± 0.003 | 1.976 ± 0.008 |
| (3782) Celle | 0.935 ± 0.003 | 1.966 ± 0.012 | 0.935 | 1.959 | 0.935 ± 0.003 | 1.963 ± 0.012 |
| (3782) Celle | 0.934 ± 0.003 | 1.974 ± 0.009 | 0.933 | 1.963 | 0.934 ± 0.003 | 1.968 ± 0.009 |
| (3782) Celle | 0.936 ± 0.003 | 1.956 ± 0.006 | 0.933 | 1.972 | 0.935 ± 0.003 | 1.963 ± 0.008 |
| (3782) Celle (Moskovitz) | 0.939 ± 0.003 | 1.930 ± 0.010 | -- | -- | 0.939 ± 0.003 | 1.930 ± 0.010 |
| (3782) Celle (Moskovitz) | 0.944 ± 0.003 | 1.952 ± 0.004 | -- | -- | 0.944 ± 0.003 | 1.952 ± 0.004 |
| (3849) Incidentia | 0.951 ± 0.003 | 1.992 ± 0.007 | 0.947 | 2.006 | 0.949 ± 0.003 | 1.999 ± 0.007 |
| (3849) Incidentia | 0.951 ± 0.003 | 1.987 ± 0.009 | 0.949 | 2.003 | 0.950 ± 0.003 | 1.995 ± 0.009 |
| (4055) Magellan | 0.929 ± 0.003 | 1.935 ± 0.004 | 0.936 | 1.950 | 0.933 ± 0.003 | 1.943 ± 0.008 |
| (4900) Maymelou | 0.942 ± 0.003 | 1.975 ± 0.008 | 0.940 | 1.984 | 0.941 ± 0.003 | 1.980 ± 0.008 |
| (5754) 1992 $FR_2$ | 0.934 ± 0.003 | 1.951 ± 0.005 | 0.934 | 1.964 | 0.934 ± 0.003 | 1.958 ± 0.007 |
| (5952) Davemonet | 0.941 ± 0.003 | 1.978 ± 0.007 | 0.940 | 1.987 | 0.941 ± 0.003 | 1.983 ± 0.007 |
| (7302) 1993 CQ | 0.963 ± 0.005 | 1.930 ± 0.012 | 0.964 | 1.940 | 0.964 ± 0.003 | 1.935 ± 0.012 |
| (8271) Imai | 0.937 ± 0.003 | 1.944 ± 0.008 | 0.935 | 1.953 | 0.936 ± 0.003 | 1.949 ± 0.008 |
| (8271) Imai | 0.938 ± 0.003 | 1.947 ± 0.006 | 0.936 | 1.950 | 0.937 ± 0.003 | 1.949 ± 0.006 |
| (9064) Johndavies | 0.953 ± 0.003 | 2.005 ± 0.013 | 0.947 | 2.035 | 0.950 ± 0.003 | 2.020 ± 0.015 |
| (9223) Leifandersson | 0.944 ± 0.003 | 1.980 ± 0.005 | 0.943 | 2.004 | 0.944 ± 0.003 | 1.992 ± 0.012 |
| (9223) Leifandersson | 0.950 ± 0.003 | 1.970 ± 0.003 | 0.949 | 1.989 | 0.950 ± 0.003 | 1.980 ± 0.010 |
| (9368) Esashi | 0.956 ± 0.003 | 2.013 ± 0.011 | 0.954 | 2.014 | 0.955 ± 0.003 | 2.014 ± 0.011 |
| (9531) Jean-Luc | 0.949 ± 0.004 | 1.953 ± 0.008 | 0.942 | 1.972 | 0.946 ± 0.004 | 1.963 ± 0.010 |
| (10537) 1991 $RY_{16}$ | 0.959 ± 0.003 | 1.945 ± 0.020 | 0.984 | 1.981 | 0.976 ± 0.009 | 1.961 ± 0.020 |
| (10537) 1991 RY16 (Moskovitz) | 0.960 ± 0.002 | 1.933 ± 0.005 | -- | -- | 0.960 ± 0.002 | 1.933 ± 0.005 |
| (10666) Feldberg | 0.939 ± 0.003 | 1.955 ± 0.007 | 0.937 | 1.963 | 0.938 ± 0.003 | 1.959 ± 0.007 |
| (10666) Feldberg | 0.941 ± 0.003 | 1.956 ± 0.007 | 0.937 | 1.970 | 0.939 ± 0.003 | 1.963 ± 0.007 |
| (11341) Babbage | 0.947 ± 0.003 | 2.002 ± 0.008 | 0.944 | 2.009 | 0.946 ± 0.003 | 2.006 ± 0.008 |
| (11699) 1998 $FL_{105}$ | -- | -- | -- | -- | -- | -- |
| (12073) Larimer | 0.927 ± 0.003 | 1.940 ± 0.010 | 0.925 | 1.969 | 0.926 ± 0.003 | 1.955 ± 0.015 |
| (14390) 1990 $QP_{10}$ | 0.962 ± 0.003 | 1.984 ± 0.018 | 0.969 | 1.977 | 0.966 ± 0.004 | 1.981 ± 0.018 |
| (15630) Disanti | 0.939 ± 0.003 | 1.988 ± 0.007 | 0.936 | 1.991 | 0.938 ± 0.003 | 1.990 ± 0.007 |
| (16703) 1995 $ER_7$ | 0.943 ± 0.003 | 1.981 ± 0.020 | 0.941 | 1.987 | 0.942 ± 0.003 | 1.984 ± 0.020 |
| (17480) 1991 $PE_{10}$ | 0.948 ± 0.006 | -- | 0.940 | 1.941 | 0.944 ± 0.006 | 1.941 |
| (17035) Velichko | 0.933 ± 0.003 | 1.925 ± 0.010 | 0.933 | 1.946 | 0.933 ± 0.003 | 1.936 ± 0.011 |
| (19165) Nariyuki | 0.945 ± 0.003 | 1.983 ± 0.007 | 0.941 | 1.988 | 0.943 ± 0.003 | 1.986 ± 0.007 |
| (19738) Callinger | 0.942 ± 0.003 | 2.006 ± 0.013 | 0.938 | 2.051 | 0.940 ± 0.003 | 2.029 ± 0.023 |
| (20171) 1996 $WC_2$ | 0.939 ± 0.003 | 1.978 ± 0.018 | 0.937 | 1.975 | 0.938 ± 0.003 | 1.977 ± 0.018' |
| (24014) 1999 $RB_{118}$ | 0.998 ± 0.005 | -- | 1.037 | -- | 1.018 ± 0.020 | -- |
| (25849) 2000 $ET_{107}$ | 0.913 ± 0.015 | -- | -- | -- | 0.913 ± 0.015 | -- |
| (26417) Michaelgord | -- | -- | -- | -- | -- | -- |
| (27025) 1998 $QY_{77}$ | 0.922 ± 0.003 | 1.906 ± 0.017 | 0.926 | 1.892 | 0.924 ± 0.003 | 1.899 ± 0.017 |
| (34698) 2001 $OD_{22}$ | 0.928 ± 0.003 | 1.910 ± 0.010 | 0.923 | 1.940 | 0.926 ± 0.003 | 1.925 ± 0.015 |
| (34698) 2001 $OD_{22}$ | 0.933 ± 0.004 | 1.907 ± 0.008 | 0.928 | 1.921 | 0.931 ± 0.004 | 1.914 ± 0.010 |
| (36118) 1999 $RE_{135}$ | 0.955 ± 0.003 | 1.946 ± 0.014 | 0.954 | 1.969 | 0.955 ± 0.003 | 1.958 ± 0.012 |
| (36118) 1999 $RE_{135}$ | 0.950 ± 0.003 | 1.950 ± 0.011 | 0.948 | 1.952 | 0.949 ± 0.003 | 1.951 ± 0.011 |

Table 4. Band Area Ratios and average surface pyroxene chemistries for pyroxene-bearing asteroids in this paper.

| Asteroid | Gaffey et al. (2002) Avg. pyroxene chemistry | Burbine et al. (2009) Avg. pyroxene chemistry | Overall average surface pyroxene chemistry | BAR |
|---|---|---|---|---|
| (2168) Swope | $Wo_{15}Fs_{40}$ | $Wo_{11}Fs_{46}$ | $Wo_{13}Fs_{43}$ | 2.402 ± 0.245 |
| (2168) Swope | $Wo_{15}Fs_{41}$ | $Wo_{12}Fs_{48}$ | $Wo_{14}Fs_{44}$ | 2.253 ± 0.107 |
| (2168) Swope | $Wo_{13}Fs_{40}$ | $Wo_{10}Fs_{44}$ | $Wo_{12}Fs_{42}$ | 2.406 ± 0.146 |
| (3715) Stohl | $Wo_{14}Fs_{41}$ | $Wo_{12}Fs_{47}$ | $Wo_{13}Fs_{44}$ | 2.414 ± 0.101 |
| (3715) Stohl | $Wo_{13}Fs_{41}$ | $Wo_{11}Fs_{45}$ | $Wo_{12}Fs_{43}$ | 2.252 ± 0.132 |
| (3782) Celle | $Wo_{11}Fs_{43}$ | $Wo_{10}Fs_{41}$ | $Wo_{11}Fs_{42}$ | 2.117 ± 0.020 |
| (3782) Celle | $Wo_{10}Fs_{44}$ | $Wo_{9}Fs_{42}$ | $Wo_{10}Fs_{43}$ | 2.402 ± 0.240 |
| (3782) Celle | $Wo_{11}Fs_{43}$ | $Wo_{10}Fs_{41}$ | $Wo_{11}Fs_{42}$ | 2.318 ± 0.034 |
| (3782) Celle (Moskovitz) | $Wo_{12}Fs_{38}$ | $Wo_{8}Fs_{40}$ | $Wo_{10}Fs_{39}$ | 3.350 ± 0.062 |
| (3782) Celle (Moskovitz) | $Wo_{15}Fs_{40}$ | $Wo_{10}Fs_{45}$ | $Wo_{13}Fs_{43}$ | 3.040 ± 0.076 |
| (3849) Incidentia | $Wo_{17}Fs_{42}$ | $Wo_{13}Fs_{52}$ | $Wo_{15}Fs_{47}$ | 2.029 ± 0.117 |
| (3849) Incidentia | $Wo_{17}Fs_{42}$ | $Wo_{14}Fs_{52}$ | $Wo_{16}Fs_{47}$ | 2.079 ± 0.088 |
| (4055) Magellan | $Wo_{10}Fs_{37}$ | $Wo_{8}Fs_{39}$ | $Wo_{9}Fs_{38}$ | 2.135 ± 0.014 |
| (4900) Maymelou | $Wo_{13}Fs_{41}$ | $Wo_{11}Fs_{46}$ | $Wo_{12}Fs_{44}$ | 2.450 ± 0.117 |
| (5754) 1992 $FR_{2}$ | $Wo_{10}Fs_{41}$ | $Wo_{9}Fs_{41}$ | $Wo_{10}Fs_{41}$ | 2.346 ± 0.091 |
| (5952) Davemonet | $Wo_{13}Fs_{41}$ | $Wo_{11}Fs_{47}$ | $Wo_{12}Fs_{44}$ | 2.491 ± 0.309 |
| (7302) 1993 CQ | -- | -- | -- | 0.859 ± 0.140 |
| (8271) Imai | $Wo_{11}Fs_{39}$ | $Wo_{9}Fs_{41}$ | $Wo_{10}Fs_{40}$ | 2.558 ± 0.088 |
| (8271) Imai | $Wo_{12}Fs_{39}$ | $Wo_{9}Fs_{41}$ | $Wo_{11}Fs_{40}$ | 2.246 ± 0.059 |
| (9064) Johndavies | $Wo_{17}Fs_{43}$ | $Wo_{15}Fs_{55}$ | $Wo_{16}Fs_{49}$ | 2.122 ± 0.122 |
| (9223) Leifandersson | $Wo_{15}Fs_{42}$ | $Wo_{12}Fs_{49}$ | $Wo_{14}Fs_{46}$ | 2.205 ± 0.197 |
| (9223) Leifandersson | $Wo_{17}Fs_{41}$ | $Wo_{13}Fs_{51}$ | $Wo_{15}Fs_{46}$ | 2.075 ± 0.075 |
| (9368) Esashi | $Wo_{19}Fs_{43}$ | $Wo_{16}Fs_{57}$ | $Wo_{18}Fs_{49}$ | 1.990 ± 0.017 |
| (9531) Jean-Luc | $Wo_{15}Fs_{40}$ | $Wo_{12}Fs_{47}$ | $Wo_{14}Fs_{44}$ | 2.025 ± 0.031 |
| (10537) 1991 $RY_{16}$ | -- | -- | -- | 0.992 ± 0.017 |
| (10537) 1991 RY16 (Moskovitz) | -- | -- | -- | 1.020 ± 0.033 |
| (10666) Feldberg | $Wo_{12}Fs_{40}$ | $Wo_{10}Fs_{43}$ | $Wo_{11}Fs_{42}$ | 2.205 ± 0.105 |
| (10666) Feldberg | $Wo_{12}Fs_{40}$ | $Wo_{10}Fs_{44}$ | $Wo_{11}Fs_{42}$ | 2.137 ± 0.105 |
| (11341) Babbage | $Wo_{15}Fs_{43}$ | $Wo_{13}Fs_{52}$ | $Wo_{14}Fs_{48}$ | 2.137 ± 0.089 |
| (11699) 1998 $FL_{105}$ | -- | -- | -- | -- |
| (12073) Larimer | $Wo_{7}Fs_{41}$ | $Wo_{7}Fs_{36}$ | $Wo_{7}Fs_{39}$ | 2.897 ± 0.163 |
| (14390) 1990 $QP_{10}$ | -- | -- | -- | 0.895 ± 0.015 |
| (15630) Disanti | $Wo_{12}Fs_{42}$ | $Wo_{11}Fs_{46}$ | $Wo_{12}Fs_{44}$ | 2.468 ± 0.092 |
| (16703) 1995 $ER_{7}$ | $Wo_{14}Fs_{41}$ | $Wo_{12}Fs_{47}$ | $Wo_{13}Fs_{44}$ | 2.715 ± 0.465 |
| (17480) 1991 $PE_{10}$ | -- | -- | -- | 0.397 ± NA |
| (17035) Velichko | $Wo_{10}Fs_{36}$ | $Wo_{8}Fs_{38}$ | $Wo_{9}Fs_{37}$ | 2.731 ± 0.401 |
| (19165) Nariyuki | $Wo_{14}Fs_{41}$ | $Wo_{12}Fs_{48}$ | $Wo_{13}Fs_{45}$ | 2.144 ± 0.128 |
| (19738) Callinger | $Wo_{13}Fs_{44}$ | $Wo_{13}Fs_{51}$ | $Wo_{13}Fs_{48}$ | 2.286 ± 0.031 |
| (20171) 1996 $WC_{2}$ | $Wo_{12}Fs_{41}$ | $Wo_{11}Fs_{45}$ | $Wo_{12}Fs_{43}$ | 3.069 ± 0.459 |
| (24014) 1999 $RB_{118}$ | -- | -- | -- | 1.267 ± 0.009 |
| (25849) 2000 $ET_{107}$ | -- | -- | -- | 1.783 ± 0.135 |
| (26417) Michaelgord | -- | -- | -- | -- |
| (27025) 1998 $QY_{77}$ | $Wo_{6}Fs_{26}$ | $Wo_{4}Fs_{30}$ | $Wo_{5}Fs_{28}$ | 2.530 ± 0.050 |
| (34698) 2001 $OD_{22}$ | $Wo_{7}Fs_{33}$ | $Wo_{6}Fs_{33}$ | $Wo_{7}Fs_{33}$ | 1.742 ± 0.192 |
| (34698) 2001 $OD_{22}$ | $Wo_{9}Fs_{30}$ | $Wo_{7}Fs_{35}$ | $Wo_{8}Fs_{33}$ | 1.363 ± 0.205 |
| (36118) 1999 $RE_{135}$ | $Wo_{19}Fs_{40}$ | $Wo_{15}Fs_{39}$ | $Wo_{17}Fs_{40}$ | 2.251 ± 0.581 |
| (36118) 1999 $RE_{135}$ | $Wo_{13}Fs_{52}$ | $Wo_{12}Fs_{46}$ | $Wo_{13}Fs_{49}$ | 1.539 ± 0.109 |

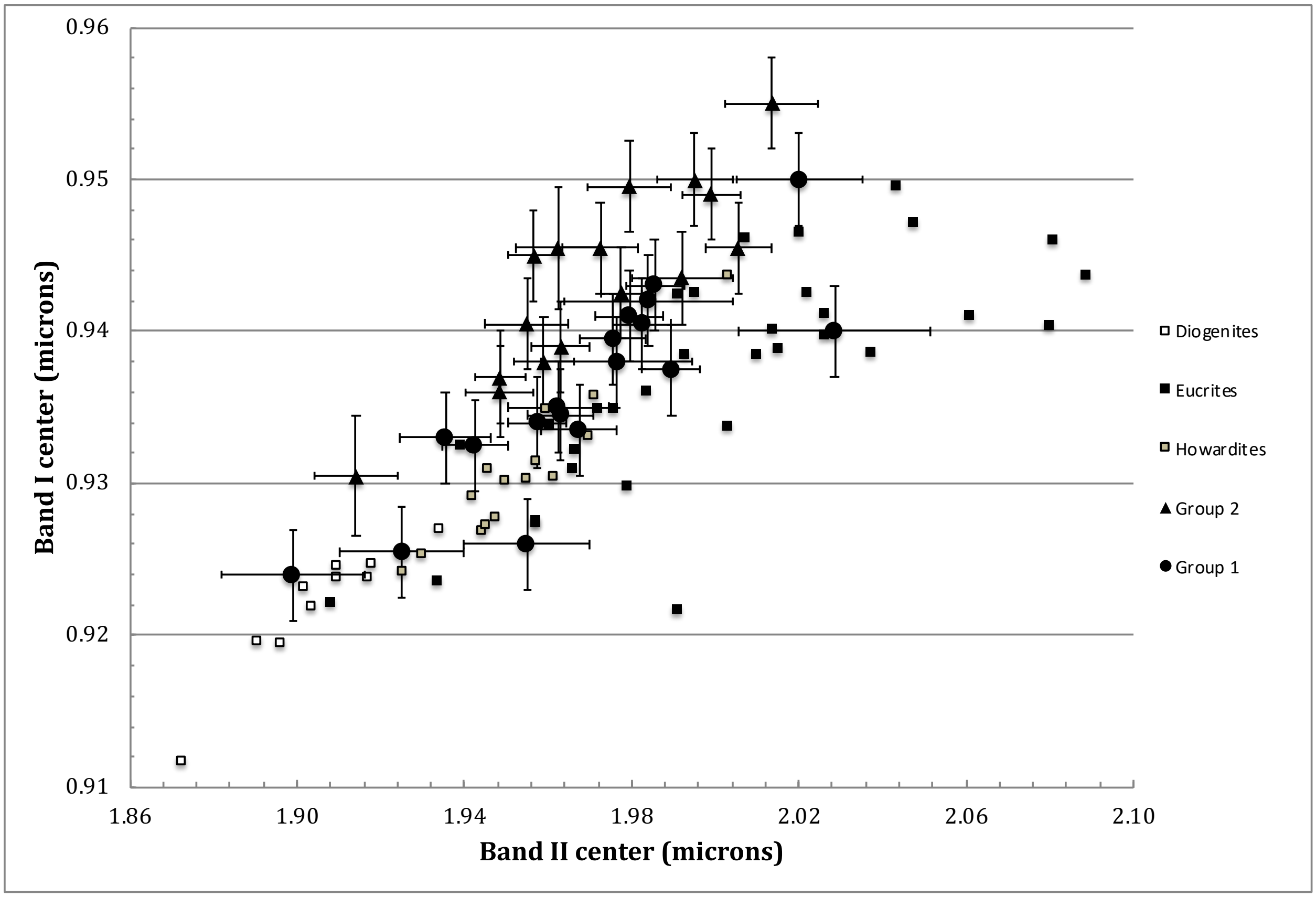

Figure 3. Band I vs. Band II center plot for the inner-belt $V_p$ asteroids in this work. The 16 Group 1 asteroids plot directly on the HED meteorite data in the plot while the eight Group 2 asteroids plot slightly above the data for the HED meteorites. Both groups of asteroids exhibit average surface pyroxene chemistries consistent with the HED meteorites.

Group 1. Fifteen of the $V_p$ asteroids in our sample exhibit spectral, band parameter, and mineralogical evidence that these asteroids have a basaltic surface composition. Spectrally, Figure 2 displays the NIR spectra of these asteroids that exhibit the two deep pyroxene absorptions at ~0.9- and ~1.9-µm with continuum-removed Band I and Band II depths that range from 35-52% and 32-55%, respectively. Plotting the continuum-removed Band I and Band II centers in Figure 3 places all the inner-belt $V_p$ asteroids on the plot with the data for the HED meteorite suite of howardites, diogenites, and eucrites. Figure 4 is a band-band plot for the outer-belt $V_p$ asteroids in this work. Figure 5 shows that all Group 1 asteroids plot on the Band I vs. BAR plot in the rectangular region designated for basaltic achondrites or with larger BAR values that are beyond the bounds of that region.

Using the band center data in the laboratory pyroxene calibration of Gaffey et al. (2002) and the laboratory HED meteorite calibration of Burbine et al. (2009), and averaging the results, each asteroid was tested to determine if it had an association with a specific type of HED meteorite. Three $V_p$ asteroids [(9064) Johndavies, (11341) Babbage, (19738) Callinger] produced average surface pyroxene chemistries consistent with the eucrites.

Only one asteroid, (27025) 1998 $QY_{77}$, has a surface mineralogy consistent with the diogenites. Three asteroids [(4055) Magellan, (5754) 1992 $FR_2$, (12073) Larimer] are consistent with a howardite interpretation. One asteroid, (17035) Velichko, has a likely howardite composition that may be enriched in a diogenite component while seven asteroids [(3782) Celle, (4900) Maymelou, (5952) Davemonet, (15630) Disanti, (16703) 1995 $ER_7$, (19165) Nariyuki, (20171) 1996 $WC_2$) are consistent with a howardite interpretation with a possible eucrite enrichment.

Group 2. This group of asteroids is very similar to the Group 1 asteroids with the sole difference being that these asteroids plot just above the Band I vs. Band II HED meteorite data in Figure 3A. This slight offset in position in Figure 3A, however, does not affect the resulting interpretations for these asteroids. The HED data in Figure 3A does not encompass the entire range of measured HED band center data, which extend out to ~0.98-μm for Band I (Figure 4, Cloutis et al., 2013). The asteroid Band I positions may also suggest a slight enrichment in an olivine or high-Ca clinopyroxene component to the average surface composition (Hardersen et al., 2014), but not enough of an enrichment to lead to a significant non-basaltic surface composition.

For this group, three asteroids exhibit average surface pyroxene chemistries consistent with eucrites [(3849) Incidentia, (9223) Leifandersson, (9368) Esashi]; one asteroid, (8271) Imai, is consistent with a howardite interpretation; and four asteroids are consistent with a howardite interpretation with a possibly enhanced eucrite component [(2168) Swope, (3715) Stohl, (9531) Jean-Luc, (10666) Feldberg].

Group 3. Only two inner-belt $V_p$ asteroids are completely inconsistent with a basaltic surface composition: (11699) 1998 $FL_{105}$ and (26417) Michaelgord. Both asteroids display a featureless NIR spectrum in Figure 2 with (11699) 1998 $FL_{105}$ displaying a relatively steep, spectrally red slope with increasing reflectance from ~0.7- to 1.8-μm, followed by a flattening of the spectrum beyond 1.8-μm. The weak absorptions at ~1.4- and ~1.9-μm in the NIR spectrum of (11699) 1998 $FL_{105}$ are likely telluric. (26417) Michaelgord displays a much shallower increase in reflectance across the same spectral range but is also spectrally featureless. **However, Jasmin et al. (2013) report a NIR spectrum and analysis for (11699) 1998 $FL_{105}$ that is consistent with a basaltic surface composition. Verification of the basaltic nature of (11699) 1998 $FL_{105}$ will require additional observations, but our results should be taken as provisional pending verification of the presence of rotational variations in the asteroid's NIR spectra.**

Both asteroids are members of the Vesta dynamical family (Zappala et al., 1995; Nesvorny, 2015) and are classified as $V_p$ asteroids with no classification variability (Carvano et al., 2010). (11699) 1998 $FL_{105}$ has a $V_p$ classification based on one SDSS observation, has a classification probability score of 89%, and a SDSS *z'* filter mean error (0.008) that is below the average error at the *z'* filter for all SDSS *z'* filter observations for the asteroids in this work (0.015). The NEOWISE geometric albedo, $p_v$, for this asteroid is $0.245 \pm 0.034$ (Masiero et al., 2011). Additional NIR spectra of (11699) 1998 $FL_{105}$ should be obtained to confirm our results.

For (26417) Michaelgord, there are two SDSS observations, both with $V_p$ classifications, probability scores of 11% and 64%, and with both observations being flagged as 'BAD'. From Hasselman et al. (2012), a 'BAD' flag defines at least one derived magnitude having an uncertainty above the 3rd quartile for the observation associated with the assigned classification. The mean SDSS *z'* filter errors for the two observations for (26417) Michaelgord are 0.024 and 0.014, which are near, or larger than, the mean SDSS *z'* filter error for all the asteroids in this work (0.015). The NEOWISE geometric albedo, $p_v$, for (26417) Michaelgord is 0.472 ± 0.199. Additional NIR spectra of this asteroid should also be obtained to confirm our findings.

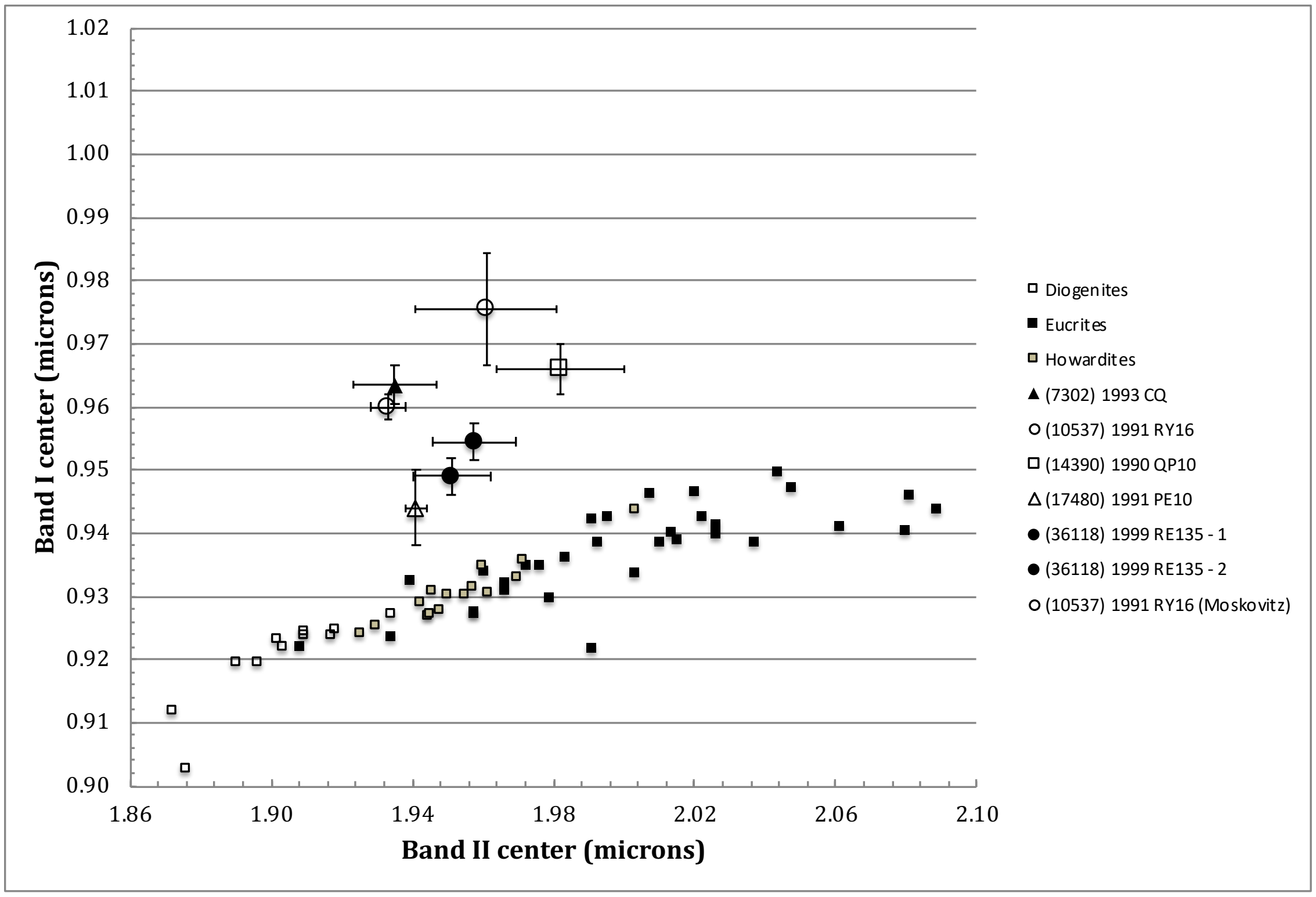

Figure 4. Band I vs. Band II plot for the outer-belt basaltic asteroid candidates in this work. Also included is a plot of band center derived data from a NIR spectrum of (10537) 1991 $RY_{16}$ from Moskovitz et al. (2008).

Outer-belt $V_p$ Asteroids. Eight of the asteroids in this work have semimajor axes that place them beyond the mean motion resonance at 2.5 AU and are considered "outer main belt" asteroids unlikely to be fragments from (4) Vesta (Lazzaro et al., 2000). Their semimajor axes are quite disparate and range from ~2.56 AU for (24014) 1999 $RB_{118}$ to ~3.24 AU for (14390) 1990 $QP_{10}$. For reference, the only previously identified basaltic asteroid in the outer main belt, (1459) Magnya, has a semimajor axis of ~3.14 AU. Three asteroids have been classified into different families: (17480) 1991 $PE_{10}$ is a member of the Ceres dynamical family (Zappala et al., 1995), (25849) 2000 $ET_{107}$ is a member of the Maria or Maria/Renate family (Nesvorny, 2015; Mothe-Diniz et al., 2012), and (34698) 2001 $OD_{22}$ is a member of the Alauda family (Nesvorny, 2015) and a member of the C4 high-inclination asteroid family (*i* ~23.2°, Gil-Hutton, 2007). The remaining outer-belt $V_p$ asteroids, (7302) 1993 CQ, (10537) 1991 $RY_{16}$, (14390) 1990 $QP_{10}$, (24014) 1999 $RB_{118}$, and (36118) 1999 $RE_{135}$ have not been assigned to any family thus far. This evidence suggests that these asteroids are probably unrelated to each other and derive from different parent bodies.

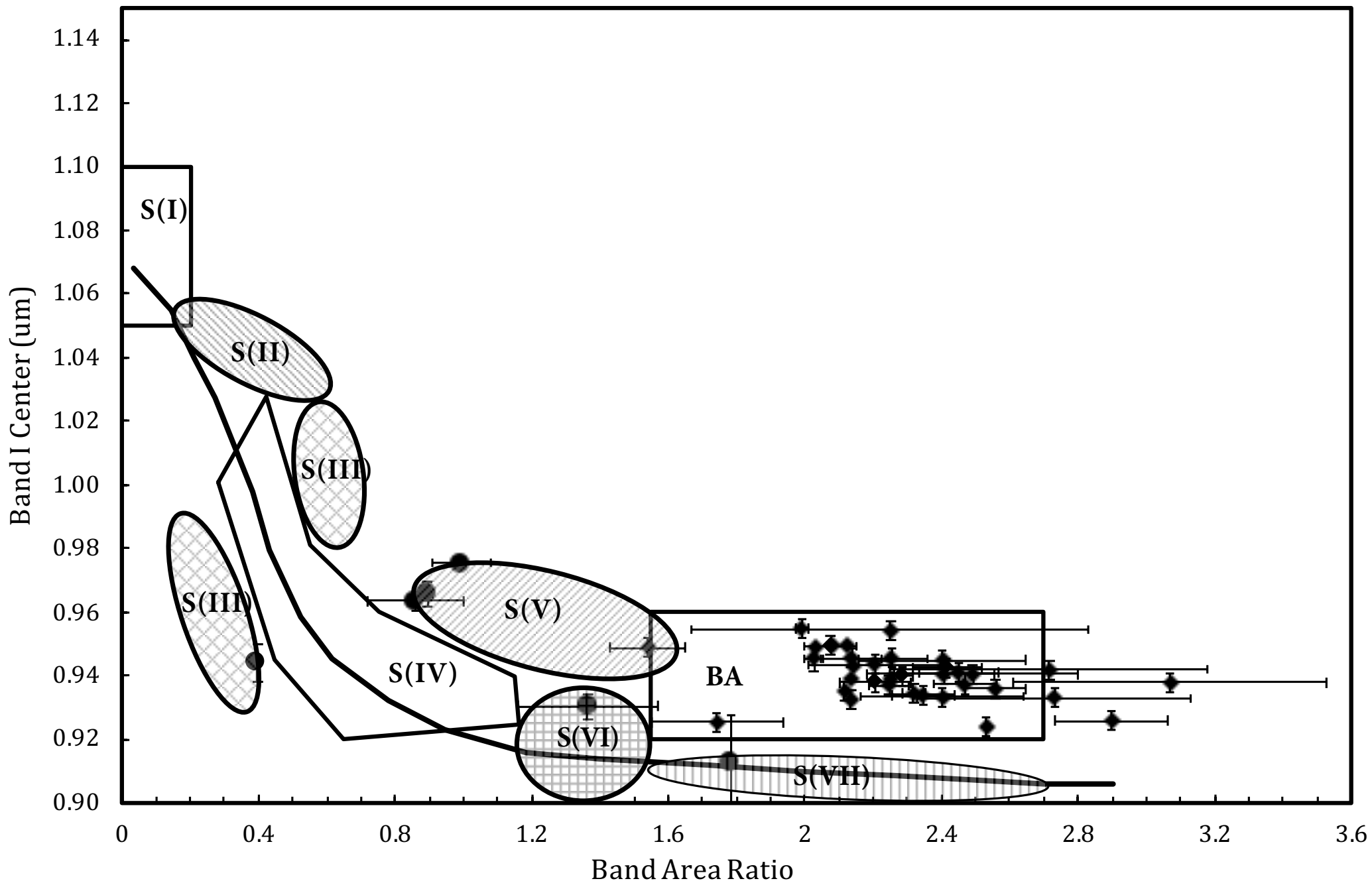

Figure 5. Band I center vs. Band Area Ratio (BAR) plot for the $V_p$ asteroids in this work. The rectangular region encloses band parameter space for basaltic achondrites (BA) and represent asteroids with surfaces that are dominated by pyroxene group minerals and are consistent with a basaltic composition. The regions labels S(I) through S(VII) taken from Gaffey et al. (1993).

Six of these asteroids are classified as $V_p$ asteroids, but two are classified as SV asteroids [(17480) 1991 $PE_{10}$, (25849) 2000 $ET_{107}$] (Hasselmann et al., 2012). Both latter asteroids are based on one SDSS observation and both have a 'BAD' flag associated with the taxonomic classification. (17480) 1991 $PE_{10}$ has a taxonomic probability score of 10% and (25849) 2000 $ET_{107}$ has a taxonomic probability score of 8% (Hasselmann et al., 2012). (10537) 1991 $RY_{16}$ has taxonomic probability scores of 47% and 89% based on two SDSS observations, but no 'BAD' flag designations (Hasselmann et al., 2012). (14390) 1990 $QP_{10}$ has a $V_p$ classification based on two SDSS observations with taxonomic probability scores of 8% and 6%, respectively, both with 'BAD' flag markers (Hasselmann et al., 2012). The $V_p$ classification for (34698) 2001 $OD_{22}$ is based on one SDSS observation with a taxonomic probability score of 14% and marked as 'BAD' (Hasselmann et al., 2012). Finally, the $V_p$ classification for (36118) 1999 $RE_{135}$ is based on one SDSS observation, has a taxonomic probability score of 11% and a 'BAD' flag rating (Hasselmann et al., 2012).

Visually, the NIR spectra of most outer-belt basaltic asteroid candidates are markedly different from the NIR spectra of asteroids that are consistent with a basaltic surface composition. The absorption features of the ~0.9- and ~1.9-μm absorptions for these asteroids, if they are present, exhibit different spectral band shapes and depths that indicate differences from primary pyroxene absorption features. While the Band I and Band II continuum-removed absorption band depths for likely basaltic asteroids ranges from 31% to 52% and 31% to 46%, respectively, the band depths for the outer-belt $V_p$ asteroids range from 11% to 53% and 9% to 30%, respectively. A Band II depth for (17480) 1991 $PE_{10}$ could not be measured due to the weakness of the feature and the noisiness of that region of the spectrum. [1]

(34698) 2001 $OD_{22}$ and (36118) 1999 $RE_{135}$ are the only two outer-belt $V_p$ asteroids in our sample that are both spectrally similar to, and consistent with, a basaltic surface composition. For (34698) 2001 $OD_{22}$, Band I and II centers plot on or slightly above the plot in Figure 3A and the derived average surface pyroxene chemistries suggest that this asteroid has a howardite-like composition ($Wo_8Fs_{33}$) possibly enriched with a diogenite component. While the BAR (1.742) for one NIR spectrum is consistent with a basaltic achondrite analog, the other BAR (1.363) is more consistent with the S-VI asteroids (Gaffey et al., 1993). Potential meteorite analogs for the S-VI group include lodranites, winonanites/IAB irons, and siderophyres (Gaffey et al., 1993). The BAR variations suggest the intriguing possibility of surface compositional variations with rotation but require further observations for confirmation.

For (36118) 1999 $RE_{135}$, plots of Band I and Band II for both nights of data are marginally above the HED trend line in Figure 4. BAR values ranging from ~1.5 to ~2.25, along with the range of average surface pyroxene chemistries (~$Fs_{40\text{-}49}$), are consistent with a howardite or eucrite interpretation. As discussed with the Group 2 asteroids, band data that plots only marginally above the HED trend line may have some additional olivine or clinopyroxene within the overall surface composition, but can still be consistent with an HED meteorite analog.

Five of the asteroids [(7302), (10537), (14390), (17480), (24014)] exhibit Band I and Band II absorptions, but only one asteroid, (17480) 1991 $PE_{10}$, has a potential meteorite analog. Band parameters for (17480) 1991 $PE_{10}$ suggest a possible H/L-chondrite analog from the Dunn et al. (2010) ordinary chondrite calibration; however, the Gaffey et al. (2002) ordinary chondrite calibration fails for all three ordinary chondrite types. The remaining asteroids with Band I and II absorptions fail both the Gaffey et al. (2002) and Dunn et al. (2010) ordinary chondrite calibration tests. Regardless, it is very likely that these asteroids exhibit some combination of mafic silicate minerals on their surfaces (i.e., olivine, low- and/or high-Ca pyroxene) based on the presence of these absorption features and their Band I and Band II center values. Those asteroids that plot above the olivine-orthopyroxene trend line in Figure 4, with the exception of (36118) 1999 $RE_{135}$, have relatively large amounts of olivine or clinopyroxene presents on their surfaces, which excludes them from classification as likely basaltic asteroids.

[1] Note that the Band II depth of ~22% for (31414) Rotarysusa in Hardersen et al. (2015) is incorrect and is more accurately shown to be ~42%.

**Jasmin et al. (2013) reported a NIR spectrum for (7302) 1993 CQ that is very similar to our spectrum in Figure 2. Their reported band parameters (Band I: 0.958 μm; Band II: 1.914 μm; BAR: 0.566) are broadly consistent with our results, although our reported band centers and BAR values are marginally larger. Jasmin et al. (2013) suggested an ordinary chondrite meteorite analog for (7302) 1993 CQ. Our two results are consistent in that (7302) 1993 CQ is inconsistent with a basaltic surface composition.**

(25849) 2000 $ET_{107}$ is spectrally unique among these $V_p$ asteroids as only weak Band I and (possibly) Band II absorption features are present in the NIR spectrum. The Band I center at ~0.91-μm is consistent with a surface dominated by low-Fe orthopyroxenes and possibly metal due to the overall redness of the spectrum. These spectral and band parameter characteristics are consistent with many of the M-/X-type asteroids analyzed by Hardersen et al. (2011). Interpretations suggested by Hardersen et al. (2011) for asteroids with similar spectral features include residual mantle material overlying disrupted Fe cores of differentiated parent bodies, analogs to the CB-/CH-chondrites, mesosiderites, and silicate-bearing iron meteorites.

<u>Investigating $V_p$ taxonomic misclassifications</u>. To further attempt to understand the reasons for the apparent misclassifications of some of the $V_p$ asteroids in our sample, we compared the SDSS *i'* and *z'* filter wavelength positions (0.763- and 0.913-μm, respectively: Fukugita et al., 1996; Carvano et al., 2010) on the NIR spectra of the asteroids from this work displaying the ~0.9-μm absorption feature. For the $V_p$ asteroids identified as basaltic, the *i'* filter wavelength position was uniformly positioned at or near the short-wavelength local reflectance maximum of the ~0.9-μm absorption. The SDSS *z'* filter wavelength position was also uniformly positioned at or near the minimum absorption of the ~0.9-μm feature.

For the asteroids in this work identified as non-basaltic and having a ~0.9-μm absorption, the SDSS *i'* and *z'* filter wavelength positions also reasonably mirrored the results for the basaltic asteroids above. Exceptions include (14390) 1990 $QP_{10}$ where the absorption extends to shorter wavelengths than the *i'* filter wavelength position and (24014) 1999 $RB_{118}$ where the SDSS *z'* filter wavelength position is somewhat shortward of the band minimum position.

Overall, the choice of central wavelength positions for the SDSS *i'* and *z'* filters seem very well-selected to capture the short-wavelength local reflectance band maximum and band minimum for asteroids with a primarily surface pyroxene mineral signature. Asteroids with weaker Band I depths and with different surface mineral abundances are also captured in the $V_p$ taxonomy with this method, which may be due to larger than average SDSS filter noise and errors, as well as the relatively large (*i*-*r*) and (*z*-*i*) color gradient ranges for the $V_p$ asteroids. The cause of the apparent misclassifications for (11699) 1998 $FL_{105}$ and (26417) Michaelgord are unknown but should be pursued with additional NIR spectral observations.

Conclusions. Based on the results from this work and the work of Hardersen et al. (2014, 2015) for 49 potentially basaltic asteroids, we can conclude the following:

- For the 41 inner-belt $V_p$ asteroids in our sample, 39 of these asteroids exhibit NIR spectra, absorption features, derived spectral band parameters (band centers, BAR, band depths), and derived average surface mineralogies that are consistent with a basaltic interpretation with likely HED meteorite analogs. This represents an ~95% success rate of the $V_p$ taxonomy in predicting a basaltic surface composition.
- The two inner-belt $V_p$ asteroids that are **potentially** not basaltic, (11699) 1998 $FL_{105}$ and (26417) Michaelgord, are spectrally featureless in the NIR. There is no evidence from Hasselmann et al. (2012) and the SDSS data (Carvano et al., 2010) of excessive noise or errors in the data used to determine the $V_p$ classification for (11699) 1998 $FL_{105}$. By comparison, (26417) Michaelgord does exhibit the presence of noisy SDSS *z'* filter data that may have contributed to the misclassification. **Jasmin et al. (2013) report a basaltic NIR spectrum and analysis for (1699) 1998 $FL_{105}$ that suggests a basaltic surface composition. Additional NIR spectra for this asteroid are necessary to confirm our result or to provide evidence for rotational spectral variations.**
- Only two outer-belt $V_p$ asteroids in our sample, (34698) 2001 $OD_{22}$ and (36118) 1999 $RE_{135}$, are suggestive of basaltic surface compositions and an HED meteorite analog. Variations in the derived BAR values for (34698) 2001 $OD_{22}$ suggest the possibility of surface compositional variability, which requires confirmation via additional observations.
- Two outer-belt asteroids with an SV classification, (17480) 1991 $PE_{10}$ and (25849) 2000 $ET_{107}$, do not have a basaltic surface composition, exhibit relatively high SDSS *z'* filter magnitude errors, and low taxonomic probability classifications. (17480) 1991 $PE_{10}$ has a potential H- or L-chondrite meteorite analog. (25849) 2000 $ET_{107}$ has a weak Band I feature suggestive of the presence of low-Fe orthopyroxene.
- The remaining four outer-belt $V_p$ asteroids display NIR absorption features of varying shapes and band depths indicative of the presence of one or more mafic silicate surface minerals and possibly metal. However, no meteorite analogs have yet been identified for these asteroids.
- Results from Leith et al. (2017) identify (10537) 1991 $RY_{16}$ and (14390) 1990 $QP_{10}$ as Unclassified Objects (Table 6) and different from the V-type asteroids in this work, which is consistent with our results.
- **Thus far, the $V_p$ taxonomy has predicted a basaltic surface composition for only two out of eight asteroids (~25%).** The sample of outer-belt asteroids generally exhibit greater classification uncertainties via their low taxonomic probability scores, the presence of 'BAD' flags indicating larger SDSS magnitude uncertainties, and generally larger SDSS *z'* filter errors compared to the average for all the asteroids in this work.

- Fourteen $V_p$ asteroids with likely basaltic surface compositions and identified HED meteorite analogs from this work and Hardersen et al. (2014, 2015) display either a low taxonomic probability score or a 'BAD' flag indication. These indications of potentially noisy or low-quality data did not translate into a non-basaltic composition for these asteroids. By comparison, there are two $V_p$ asteroids [(7302) 1993 CQ, (11699) 1998 $FL_{105}$] that were not identified with a basaltic surface composition, but also did not display any indications of low-quality or noisy data.
- Among the outer-belt basaltic asteroid candidates, seven of the eight asteroids displayed the 'BAD' flag, had low taxonomic probability scores, or relatively high SDSS *z'* filter errors. Thirteen of the 41 inner-belt basaltic asteroid candidates displayed similar indications of low-quality data, but this did not translate into non-basaltic surface composition interpretations. One reason for this is that many of the asteroids were observed multiple times with most of the observations indicating good quality data. Regardless, a smaller percentage of inner-belt asteroids in our sample of $V_p$ asteroids displayed questionable data as compared to the outer-belt asteroid sample.
- While our results suggest that many outer-belt $V_p$ asteroids may be misclassified and do not typically indicate the presence of basaltic material, the identification of two additional likely basaltic asteroids in the outer-main belt indicates that a larger basaltic population likely exists and should be characterized. Accurately characterizing this population could affect the nature, extent, and distribution of the early solar system heating event (Herbert et al., 1991; Grimm and McSween, 1993).

# Acknowledgements

The authors acknowledge and thank the leadership, staff, and telescope operators at the NASA Infrared Telescope Facility (IRTF), as well as the program officers and scientists at NASA Headquarters, for giving us the opportunity to conduct this research program and observe at the IRTF with the SpeX instrument, both on-site and remotely.

The authors thank **Tom Burbine** for the valuable and insightful comments that improved this manuscript.

PSH thanks Sean Lindsay for access to the SARA software, and guidance on operation of the program, for a portion of our data analysis work. PSH also thanks the co-authors of this paper for their contributions.

PSH also sincerely thanks everyone who has been genuinely supportive of his personal and professional journey and transition away from the Upper Midwest to the warmer, more enjoyable and productive environs of Arizona. May fortune favor the foolish.

PSH and VR work is supported by NASA Planetary Astronomy Program Grant NNX14AJ37G (PI: Hardersen).

# Appendices

None.